\documentclass[fleqn,usenatbib]{mnras}

\usepackage{newtxtext,newtxmath}
\usepackage{gensymb}
\usepackage{lscape}
\usepackage{enumitem}
\usepackage[T1]{fontenc}
\usepackage{ae,aecompl}

\newcommand{\NHanocut}{962}
\newcommand{\NHa}{953}
\newcommand{\NHaF}{714}

\usepackage{graphicx}	% Including figure files
\usepackage{amsmath}	% Advanced maths commands
\usepackage{mathtools}	% Advanced maths commands
\usepackage{ulem}       % For strikethrough text
\usepackage{caption}
\usepackage[flushleft,referable]{threeparttablex}
\usepackage{footnote}
\usepackage{multicol}
\usepackage[dvipsnames]{xcolor}

\usepackage{floatrow}
\newcommand{\Mstar}{M_{\star}}
\newcommand{\Msun}{M_{\odot}}
\newcommand{\Hyd}{{\rm H}}
\newcommand{\Ha}{\Hyd$\alpha$}
\newcommand{\Hb}{\Hyd$\beta$}
\newcommand{\Hg}{\Hyd$\gamma$}
\newcommand{\Hbe}{\Hyd\beta}
\newcommand{\OII}{[{\sc O ii}]$\lambda$3726,3728}
\newcommand{\OIII}{[{\sc O iii}]$\lambda$5007}
\newcommand{\EBV}{E(B-V)}
\newcommand{\EBVm}{$E$($B$--$V$)}
\newcommand{\Rc}{R_{\rm C}}
\newcommand{\NII}{[\textsc{N~ii}]}
\newcommand{\NIIl}{\NII$\lambda\lambda$6548,6583}

\title[Extremely low-mass star formation sequence]{The Metal Abundances across Cosmic Time ($\mathcal{MACT}$) Survey. III -- The relationship between stellar mass and star formation rate in extremely low-mass galaxies}

\author[K. Shin et al.]{Kaitlyn Shin,$^{1,2,3}$\thanks{E-mail: kshin@mit.edu}
Chun Ly,$^{4,3}$\thanks{E-mail: chunly@email.arizona.edu}
Matthew A. Malkan,$^{5}$
Sangeeta Malhotra,$^{3}$
\newauthor
Mithi de los Reyes,$^{6}$
and James E. Rhoads$^{3}$
\\
$^{1}$Department of Physics and Kavli Institute for Astrophysics and Space Research, MIT, Cambridge, MA 02139, USA\\
$^{2}$Department of Physics, Stanford University, 452 Lomita Mall, Stanford, CA 94305, USA\\
$^{3}$Astrophysics Science Division, NASA/Goddard Space Flight Center, Code 660, Greenbelt, MD 20771, USA\\
$^{4}$Steward Observatory, University of Arizona, 933 N Cherry Avenue, Tucson, AZ 85719, USA\\
$^{5}$Department of Physics and Astronomy, UCLA, 430 Portola Plaza, Los Angeles, CA 90095, USA\\
$^{6}$Department of Astronomy, California Institute of Technology, MC 249-17, 1200 E California Blvd, Pasadena, CA 91125, USA
}

\date{Accepted XXX. Received 2019 October 23.}

\pubyear{2020}

\begin{document}
\label{firstpage}
\pagerange{\pageref{firstpage}--\pageref{lastpage}}
\maketitle

\begin{abstract}
  Extragalactic studies have demonstrated there is a moderately tight ($\approx$0.3 dex) relationship between galaxy stellar mass ($\Mstar$) and star formation rate (SFR) that holds for star-forming galaxies at $\Mstar \sim 3 \times 10^8$--10$^{11}~\Msun$, i.e., the ``star formation main sequence.'' However, it has yet to be determined whether such a relationship extends to even lower mass galaxies, particularly at intermediate or higher redshifts. We present new results using observations for \NHaF\ narrowband \Ha-selected galaxies with stellar masses between $10^6$ and $10^{10}~\Msun$ (average of $10^{8.2}~\Msun$) at $z \approx$ 0.07--0.5. These galaxies have sensitive UV to near-infrared photometric measurements and optical spectroscopy. The latter allows us to correct our \Ha\ SFRs for dust attenuation using Balmer decrements. Our study reveals:
  (1) for low-SFR galaxies, our \Ha\ SFRs systematically underpredict compared to FUV measurements, consistent with other studies;
  (2) at a given stellar mass ($ \approx $10$ ^{8}~\Msun$), log(specific~SFR) evolves as~$ A \log(1+z) $ with $ A = 5.26 \pm 0.75 $, and on average, specific SFR increases with decreasing stellar mass;
  (3) the SFR--$\Mstar$ relation holds for galaxies down to $\sim$10$^6~\Msun$ ($\sim$1.5 dex below previous studies), and over lookback times of up to 5 Gyr, follows a redshift-dependent relation of $\log{({\rm SFR})} \propto \alpha \log(\Mstar/\Msun) + \beta z$ with $\alpha = 0.60 \pm 0.01$ and $\beta = 1.86 \pm 0.07$; and
  (4) the observed dispersion in the SFR--$\Mstar$ relation at low stellar masses is $\approx$0.3 dex. Accounting for survey selection effects using simulated galaxies, we estimate the true dispersion is $\approx$0.5 dex.
\end{abstract}


%%%%%%%%%%%%%%%%%%%%%%%%%%%%%%%%%%%%%%%%%%%%%%%%%%

%%%%%%%%%%%%%%%%% BODY OF PAPER %%%%%%%%%%%%%%%%%%

\section{Introduction}
\label{sec:intro}

A major goal in galaxy formation and evolution studies is to better understand the star formation history of galaxies across cosmic time. To this end, the relationship between galactic star formation rate (SFR) and stellar mass ($\Mstar$) has been extensively studied \citep[e.g.,][and references therein]{Brinchmann04,Noeske07,Dutton10,Reyes15}. The SFR measures the instantaneous rate at which a galaxy converts its gas into stars, while $\Mstar$ is an estimate for the total amount of stars formed in the galaxy. Many recent empirical studies demonstrate that there is a relatively tight ($\sigma\sim0.3$ dex) correlation between the SFRs and stellar masses of galaxies.
This positive correlation (SFR $\propto$ $\Mstar^{\alpha}$ with $\alpha\approx0.6$--1; \citealp{Rodighiero11}) has been called the galaxy ``star formation sequence'' \citep{Salim07} or the ``main sequence'' \citep{Noeske07}.
It has been seen that this relation holds at high redshifts, such that higher redshift galaxies show higher rates of star formation than local galaxies do \citep[e.g.,][]{Whitaker12b} with a peak at $z\sim2$ \citep[e.g.,][]{Gonzalez11,Madau14}.

Much of this observed relation is limited to galaxies with moderate to massive stellar masses \citep[$\gtrsim 3 \times10^{9}\,\Msun$;][]{Karim11,Whitaker12b,Reyes15}.
Recently, \cite{McGaugh17} have investigated low surface brightness galaxies using \Ha\ SFRs in the stellar mass range of $7\times10^6$--10$^{10}~\Msun$ and found that a star-formation sequence holds; however, this study was only for the local Universe.
For studies at intermediate to higher redshifts, there have been limitations.
For example, while \cite{Whitaker14} have extended the relation to
$2.5\times10^8~\Msun$ at $z\sim0.5$,
they had stacked ultraviolet (UV) and infrared measurements to derive average SFRs.
In addition, while \citet{Atek14} investigated the relation at lower stellar masses ($\gtrsim 10^7~\Msun$) over $0.3 \lesssim z \lesssim 2.3$, their higher SFR selection limits their study to only starburst galaxies.
Finally, \cite{Iyer18} also found that the SFR--$\Mstar$ relation extends to low-mass ($\sim$10$^7~\Msun$) galaxies at higher redshifts ($z\gtrsim4$) by using a statistical approach that involved reconstructing the star formation histories of high-$z$ galaxies.
Using methods such as stacking and star formation history reconstruction, and/or being limited to high SFR galaxies, make it difficult to measure the intrinsic dispersion of the SFR--$\Mstar$ relation.
The intrinsic dispersion allows us to understand the impact of stochastic events \citep[e.g.,][]{Dave12}.  A loose correlation would imply a more merger-driven and ``bursty'' star formation history, but a dispersion of 0.3 dex (often observed) in this correlation suggests that the star formation history generally follows a smooth, regulated pattern of growth (e.g., \citealp{Daddi07}; \citealp{Renzini09}). Both theoretical and observational evidence have suggested that the dispersion in the star formation sequence increases for lower mass galaxies, $\lesssim3\times10^8$ $\Msun$ \citep[e.g.,][]{Dominguez15,Emami19}.

Only recently has it become feasible to directly measure the star-formation sequence or low stellar masses, $\sim$10$^6-3\times10^8~\Msun$, and constrain measurement of its dispersion at intermediate redshifts, $z \lesssim 0.5$, without suffering from mass completeness limits, requiring average SFR measurements from stacking, and/or indirectly inferring SFR based on the star formation histories.
This low-mass study is accomplished by identifying galaxies by their nebular emission, which has been demonstrated to provide a better galaxy census than color- or mass-limited selection \citep{Ly12b}. Here in this study, we use ultra-deep narrowband imaging from the Subaru Deep Field \citep[SDF;][]{Kashikawa04} to identify \Ha\ emitting galaxies at $z=0.07$--0.5 with stellar masses as low as 10$^6~\Msun$. This study uniquely differs from previous star formation sequence studies by (1) measuring SFR with the \Ha\ luminosity \citep[sensitive to star formation on timescales of $\lesssim$5 Myr;][]{Kennicutt12} for \textit{individual} galaxies, and (2) having deep optical spectroscopy to derive dust attenuation corrections from Balmer decrements.  In addition, deep multi-band imaging from the UV to the near-infrared is available in the SDF, which enables accurate stellar mass determinations from modeling the spectral energy distribution (SED).

The paper is outlined as follows. In Section~\ref{sec:sample} we describe the SDF data as well as follow-up spectroscopic observations. In Section~\ref{sec:methods}, we describe the sample selection used to identify \Ha\ emitting galaxies at $z=0.07$--0.5, and the spectral stacking approach that we use to yield average dust attenuation corrections from Balmer decrements and \NII/\Ha\ flux ratios. In Section~\ref{sec:props}, we use these measurements to derive \Ha\ luminosities and SFRs from the NB excess fluxes with corrections (e.g., \NII\ emission, dust attenuation). We also describe how we obtain stellar masses from SED modeling.
In Section~\ref{sec:results}, we compare our \Ha\ and FUV SFRs to examine the reliability of \Ha\ as a SFR indicator in low-SFR galaxies. 
We also discuss how our \Ha\ specific SFRs evolve with redshift and depend on stellar mass.
We then construct the SFR--$\Mstar$ relation down to $10^{6}~\Msun$ using our sample and fit the relation with a multivariate linear relation between stellar mass, redshift and SFR. We discuss the amount of intrinsic dispersion present in the SFR--$\Mstar$ relation for our sample, and the selection function of our study. We also compare our results to current empirical studies and discuss our interpretation of the results. Finally, in Section~\ref{sec:concl} we summarise our work. Throughout this paper, we adopt a flat cosmology with $\Omega_M = 0.3$, $\Omega_\Lambda = 0.7$, and $H_0 = 70$ km s$^{-1}$ Mpc$^{-1}$.

\section{Sample selection and observations}
\label{sec:sample}

\subsection{The SDF NB excess emitter sample}
\noindent
The SDF is a deep galaxy survey completed on the 8.4-m Subaru Telescope and has the most sensitive optical imaging in several narrowband (NB)\footnote{Deep imaging in intermediate band filters are also available, but are not the focus of this paper.} filters, over a field-of-view of 34\arcmin\ $\times$ 27\arcmin\ \citep{Kashikawa04}. The data from SDF were acquired with the prime-focus optical imager, Suprime-Cam \citep{Miyazaki02}, between March 2001 and May 2007. Acquisition and reduction of the SDF data are extensively discussed in \citet{Kashikawa04,Kashikawa06}, \citet{Ly07,Ly12b}, and \citet{Nagao08}. The full sample of sources in this paper consists of 7,963 emission-line galaxies selected from the NB704, NB711, NB816, NB921, and/or NB973 filters \citep{Ly16}.\footnote{A number of emission-line galaxies are excess emitters in more than one NB filter (e.g., \OIII\ emission in NB704 and \Ha\ emission in NB921 at $z\approx0.4$).} The selection of NB excess emitters used the NB color excess diagram (see e.g., \citealp{Fujita03}; \citealp{Ly07}). A summary of the properties of these filters is provided in Table~\ref{tab:filt_props}.

\begin{table}
  \centering
  \caption{Summary of NB filter properties and NB excess emitter samples}
  \label{tab:filt_props}
  \begin{threeparttable}
    \begin{tabular}{lcccr}
    \hline\hline
    Filter & $\lambda_c$ & FWHM & $z_{\rm FWHM}({\rm H}\alpha)$ & $N_{\rm NB}$ \\
    (1) &               (2) &        (3) &                        (4) & (5) \\
    \hline
    NB704 & 7045 & 100 & 0.066--0.081 & 1695 \\
    NB711 & 7126 &  72 & 0.078--0.089 & 1480 \\
    NB816 & 8152 & 120 & 0.233--0.251 & 1602 \\
    NB921 & 9193 & 132 & 0.391--0.411 & 2361 \\
    NB973 & 9749 & 200 & 0.471--0.502 & 1243 \\
    \hline
  \end{tabular}
  \begin{tablenotes}
    \item \leavevmode\kern-\scriptspace\kern-\labelsep (1): NB filter. (2): Central wavelengths in Angstroms. (3): FWHM in Angstroms. (4): \Ha\ redshift corresponding to the FWHM of NB filter. (5): Size of NB excess emitter sample.
  \end{tablenotes}
  \end{threeparttable}
\end{table}

\subsection{Follow-up optical spectroscopy}
\label{sec:spec}
\noindent
MMT's Hectospec \citep{Fabricant05} and Keck-II's Deep Imaging Multi-Object Spectrograph (DEIMOS; \citealp{Faber03}) were used to target a significant fraction ($\approx$21.5\%) of these NB emitters with follow-up optical spectroscopy. These MMT/Hectospec and Keck-II/DEIMOS observations are the primary observations for the \textit{Metal Abundances across Cosmic Time} ($\mathcal{MACT}$) survey \citep{Ly16}.

MMT/Hectospec and Keck-II/DEIMOS have complementary technical capabilities. For example, while Keck-II/DEIMOS has a higher sensitivity,
it can only target $\sim$100 galaxies at a time (within a 5\arcmin\ $\times$ 17\arcmin\ field of view) and has no spectral coverage below $\sim$6000 \AA. MMT/Hectospec, however, can target up to $\sim$270 galaxies (within a 1$\degree$ field of view) with spectral coverage down to $\sim$3700 \AA.

Given these instrumental differences, MMT/Hectospec was primarily used to target lower redshift sources ($z\lesssim0.5$) and Keck-II/DEIMOS was primarily used for higher redshift sources ($z\gtrsim0.6$), with some overlap ($z \approx 0.4$--0.65) for improved spectral coverage and consistency checks on emission-line measurements. In the overlapping redshift regime, MMT targeted brighter galaxies while Keck-II targeted fainter galaxies. Cross comparisons between spectral measurements for galaxies with both MMT and Keck-II spectra suggest that the flux calibration is reliable \citep{Ly16}.

\subsubsection{MMT/Hectospec observations}
\label{sec:MMT}

\noindent
The MMT observations took place over the equivalent of three full nights on 2008 March 13, 2008 April 10--11, 2008 April 14, 2014 February 27--28, 2014 March 25, and 2014 March 28--31. The 270~mm$^{-1}$ grating was used, which provided a spectral resolution of $R \sim 1300$, a dispersion of 1.2 \AA\ pixel$^{-1}$, and spectral coverage of 3650--9200 \AA. Data were reduced using \textsc{e-specroad}, an \textsc{iraf}-based reduction pipeline.\footnote{Developed by Juan Cabanela for use outside of CfA and available online at: \url{http://astronomy.mnstate.edu/cabanela/research/ESPECROAD/}.} There were 686 galaxies reliably detected with MMT, where a reliable detection consisted of two or more emission lines detected from either spectra or NB excess imaging.

The Hectospec data were limited by the spectral cut-off of $\approx$9200 \AA, where \Ha\ is located for $z\approx0.4$ NB921 excess emitters. In addition, the flatfielding and sky subtraction were less reliable at longer wavelengths ($\gtrsim$ 8755 \AA). This reduced reliability required that we masked the spectra for 154 targets. Each spectrum was inspected manually and masked either (1) across the entire \Ha\ emission line range or (2) across the faulty section of the spectrum redward of the \Ha\ emission line.  The latter depended on the behavior of the spectra near the \Ha\ emission line wavelength.

\subsubsection{Keck-II/DEIMOS observations}
\label{sec:Keck}
\noindent
The Keck-II observations were acquired on 2004 April 23--24 (\citealp{Kashikawa06}; \citealp{Ly07}), 2008 May 1--2 and 2009 April 25--28 \citep{Kashikawa11}, 2014 May 2 and 2015 March 17/19/26 \citep{Ly16}. The spectral resolution was $R \sim$ 3600 at 8500 \AA\ with a dispersion of 0.47 \AA\ pixel$^{-1}$. Data were reduced using the Keck-II/DEIMOS \textsc{idl}-based \textsc{spec2d} pipeline\footnote{\url{https://www2.keck.hawaii.edu/inst/deimos/pipeline.html}} \citep{Cooper12}. There were 996 galaxies reliably detected with Keck.

\section{Methods}
\label{sec:methods}

\subsection{Identifying \Ha\ NB excess emitters}
\label{sec:ha_nb_emits}
\noindent
Since the primary purpose of this study is to examine the SFR--$\Mstar$ relation with \Ha\ SFRs, we focus our analysis on \Ha\ emitters selected at $z\approx0.07$--$0.08, z\approx0.24$, $z\approx0.40$, and $z\approx0.50$. These \Ha\ emitters were primarily identified through spectroscopic redshifts, where available \citep[see Sections~\ref{sec:MMT}--\ref{sec:Keck};][]{Ly16}. When spectroscopy was unavailable, a broadband (BB) color selection technique was used. These color selections were similar to those used in \citet{Ly07} and \citet{Ly12a}, with modifications based on a larger number of galaxies with spectroscopic redshift.\footnote{The NB704, NB711, and NB816 \Ha\ selections were revised from the original selections defined in \cite{Ly07} after significant spectroscopic coverage in 2014--2016 (see Section~\ref{sec:MMT}--\ref{sec:Keck}).} For \Ha\ emitters, the color selections are:\\
\noindent NB704, NB711:
\begin{flalign}
  &V -\Rc \leq 0.84(\Rc-i^{\prime}) + 0.125,{\rm~and}\\
  &V-\Rc \geq 2.50(\Rc-i^{\prime}) - 0.240.\nonumber
\end{flalign}

\noindent NB816:
\begin{flalign}
\Rc-i^{\prime} \leq 0.45,{\rm~and~} B-V \geq 2.0(\Rc-i^{\prime}) %- 0.1
\end{flalign}

\noindent NB921:
\begin{flalign}
  &B-\Rc \geq 1.46 (\Rc-i^{\prime}) + 0.58, &\\
  &B-\Rc \leq 3.0,{\rm~and~} -0.45 \leq \Rc-i^{\prime} < 0.45.\nonumber &
\end{flalign}

\noindent NB973:
\begin{flalign}
  B-\Rc &> 2.423 (\Rc-i^{\prime}) + 0.064, &\\ %0.06386
  &\,-0.4 \leq \Rc-i^{\prime} < 0.55,{\rm~and~} 0.5 < B-\Rc \leq 3.0.\nonumber &
\end{flalign}

\noindent
These color selections are illustrated in Figure~\ref{fig:color_selection}(a)--(e) where spectroscopic samples are overlaid.
From the sample of 7,963 emission-line galaxies, \NHanocut\ of them were identified as \Ha\ emitting galaxies in the NB filters. However, nine sources identified as \Ha\ NB excess emitters have been excluded from the analysis, further reducing our sample to \NHa\ galaxies. Among these nine galaxies, four galaxies were identified as active galactic nuclei (AGN) candidates, as they have \NII$\lambda6583$/\Ha\ values above 0.54 (see Figure~\ref{fig:nii_ha}), the maximum value for star-forming galaxies \citep{Kennicutt08}. One color-selected \Ha\ NB921 emitter had an excess emission in the IA598 \citep{Nagao08}, an intermediate band filter, which we suspect is due to Ly$\alpha$ based on broad-band color information (the NB921 excess is likely due to [\textsc{C~iii}]$\lambda\lambda$1907,1909).
A color-selected \Ha\ NB816 excess emitter was excluded from the analysis because it had a photometric redshift from \textsc{eazy} \citep{Brammer08} suggesting that it may be a high-$z$ galaxy at $z\approx3$.
Another galaxy was excluded because its detection was too faint for its NB excess flux to be accurately obtained. The remaining two galaxies are identified by broad-band colors as \Ha\ NB704 emitters. However, they are also emitters in the NB973 filter; the NB704--NB973 combination for these two galaxies suggests that they are likely \Hb\ NB704 and [\textsc{S~ii}]$\lambda\lambda$6717,6731 NB973 emitters at $z\approx0.45$.

Sources that were identified as \Ha\ emitters in the NB704 and/or NB711 filters were all assigned to a combined NB704+NB711 bin as they had similar redshifts, and combining the filters also increased sample sizes for stacking (see Section~\ref{sec:stack}). Sources that were initially identified as dual \Ha\ emitters in the NB704 and NB921 filters were assigned to the NB921 filter since follow-up spectroscopy indicated that these instances occurred most likely due to [\textsc{O iii}] and \Ha\ emission in the NB704 and NB921 filters, respectively. Finally, one galaxy that was identified as a dual \Ha\ emitter in the NB816 and NB921 filter was placed in the NB921 bin upon inspection of the photometric detection.\footnote{The detection with the NB816 filter had more uncertainty associated with it than the detection with the NB921 filter.}

A summary of the redshift ranges of \Ha\ emitting galaxies for each NB filter, as well as the NB filter properties, is available in Table~\ref{tab:filt_props}.
Table~\ref{tab:spec} shows more specific details for spectroscopically confirmed cases with MMT and Keck spectra.
Spectroscopically confirmed Keck detections in the NB816 filter are not included in the analysis as they do not have \Hb\ emission-line coverage.
\newcommand{\FA}{\tnotex{tn:1}}
\begin{table*}[ht]
  \centering
  \caption{Summary of MMT and Keck spectroscopy}
  \label{tab:spec}
  \begin{threeparttable}
    \centering
    \begin{tabular}{lcccccccccc}
      \hline\hline
      Filter & $N_{\rm H \alpha}$ & $N_{\rm tgt}$ & $N_{\rm spec}$ & \multicolumn{3}{c}{MMT} & & \multicolumn{3}{c}{Keck}\\
      \cline{5-7}
      \cline{9-11}
             &                  &                  &               & $N_{\rm tgt}$ & $N_{\rm spec}$ & $z_{\rm spec}$ & & $N_{\rm tgt}$ & $N_{\rm spec}$ & $z_{\rm spec}$\\
         (1) &              (2) &              (3) &           (4) &         (5)  & (6)  & (7) & & (8) & (9) & (10)\\
  \hline
NB704 &  89 &  31 &  21 &  28 & 21 & 0.068--0.078 & & \ldots & \ldots & \ldots \\
NB711 &  50 &   7 &   3 &   7 & 3 & 0.085--0.091 & & \ldots & \ldots & \ldots \\
NB816 & 191 &  63 &  49 &  61 & 49 & 0.229--0.252 & & \ldots & \ldots & \ldots \\
NB921 & 390 & 222 & 206 & 192 & 180$^{\rm a}$ & 0.389--0.413 & & 96 & 92 & 0.389--0.428 \\
NB973 & 233 &  98 &  88 &  70 & 65 & 0.453--0.498 & & 57 & 52 & 0.458--0.498 \\
  \hline
  \end{tabular}
  \begin{tablenotes}
    \item \leavevmode\kern-\scriptspace\kern-\labelsep (1): NB filter. (2): \Ha\ sample size, $N_{\rm H \alpha}$. (3): Spectroscopic targeted sample from MMT and/or Keck, $N_{\rm tgt}$. (4): Spectroscopic sample with redshift determination from MMT and/or Keck, $N_{\rm spec}$. (5)--(7): $N_{\rm tgt}$, $N_{\rm spec}$ and $z_{\rm spec}$ range for MMT. (8)--(10): $N_{\rm tgt}$, $N_{\rm spec}$ and $z_{\rm spec}$ range for Keck. Note that there are merged detections which were targeted by both MMT and Keck.
    \item[a] \label{tn:1} 154 of 180 \Ha\ NB921 emitters had their MMT \Ha\ measurements excluded (due to poor sky subtraction, flux calibration, and/or being too close to the spectral coverage limit, 9200\AA).
  \end{tablenotes}
  \end{threeparttable}
\end{table*}

\subsection{Spectral Stacking Procedure}
\label{sec:stack}

The main goal of this study is to derive dust-corrected \Ha-based SFRs from NB photometry. This measurement requires that we correct the NB photometry for \NII\ emission and dust attenuation (discussed later in Sections~\ref{sec:NII} and \ref{sec:dust}, respectively). While these corrections can be derived from measurements on individual galaxies, only 158 of the \NHa\ \Ha\ emitting galaxies have emission line measurements reliable enough\footnote{We applied a signal-to-noise cut of $\geq$5 on the \Hb\ emission line flux.} to obtain corrections.
Therefore, in order to derive average corrections for the remainder of the galaxies in our sample, we construct composite spectra from stacking the MMT and Keck spectra.

There were 318 galaxies with spectroscopic redshift determination from MMT and 144 from Keck, and 94 of those galaxies have redshift determination from joint spectroscopy.
Of the galaxies with spectroscopic redshift determination, 315 have full \Hg\ emission-line coverage from MMT and 143 have full \Hb\ emission-line coverage from Keck, with 91 having redshift determination from joint spectroscopy.
The spectra for these galaxies were initially stacked in separate subsamples of redshift ($z$) and $\Mstar$. Based on the quality of the stacks (high signal-to-noise and large numbers of spectra in each bin), finer stacking was done---a combination of $\Mstar$ and redshift (hereafter $\Mstar$--$z$; see Figure~\ref{fig:stacked_gals_mmt} for MMT, Figure~\ref{fig:stacked_gals_keck} for Keck).

For the $\Mstar$--$z$ binning, each galaxy was assigned a $z$ bin according to its corresponding narrowband filter bin (NB704+NB711, NB816, NB921, or NB973 for MMT; NB921 or NB973 for Keck). Then, depending on how many galaxies were in the $z$ bin, each galaxy was further binned into $\Mstar$ bins. At least two $\Mstar$ sub-bins and no more than five $\Mstar$ sub-bins were required. This resulted in at least nine sources per $\Mstar$--$z$ bin.

The spectral stacking procedure was as follows. First, the spectrum of each galaxy was shifted to the rest-frame and interpolated to a proper rest-frame wavelength grid. We then constructed average spectra and fitted the resulting composite spectra with Gaussian profiles. The H$\gamma$ and \Hb\ lines were fit using double Gaussian line profiles, each with one positive and one negative component, to accurately account for Balmer stellar absorption near the centre of the emission lines. The \Ha\ line was fit using a single positive Gaussian function. The [\textsc{N ii}]$\lambda\lambda$6548,6583 emission lines near the \Ha\ lines were also fit in order to correct the NB excess flux measurements for contamination from the [\textsc{N ii}] lines (see Section~\ref{sec:NII}).
In the $\Mstar$--$z$ stacks, the [\textsc{N ii}]/\Ha\ flux ratio, a metallicity indicator, increases as expected with increasing stellar mass \citep[i.e., the mass--metallicity relation;][]{Erb06}. All fits were done using the $\texttt{scipy.optimize.curve\_fit}$ routine (version 1.0.1) in Python (version 3.7.3). From these Gaussian profiles, we derive line fluxes, equivalent widths, and flux ratios for the composite spectra (provided in Tables~\ref{tab:mmt_stacked_gals} and \ref{tab:keck_stacked_gals}).

\section{Derived properties}
\label{sec:props}

This study focuses on the relationship between the \Ha\ SFR and stellar mass. These measurements are discussed in Sections~\ref{sec:SFR} and \ref{sec:SED}, respectively.

\subsection{\Ha-based SFRs}
\label{sec:SFR}
\noindent
In this study, we use \Ha\ luminosities to derive SFRs over other SFR indicators (e.g., UV+IR). There are a few reasons why the \Ha\ luminosity is more robust for our study. First, \Ha\ directly traces high mass star formation, as it is a recombination line arising from the ionizing radiation produced by short-lived OB stars. Second, for low-luminosity or low-mass galaxies, individual measurements of UV and infrared are either difficult or not feasible. Third, with \Ha\ probing a short time-scale ($\lesssim$5 Myr) for star formation, the dispersion of the SFR--$\Mstar$ relation is more accurately measured.

To obtain dust-corrected \Ha-based SFR measurements, we use emission-line fluxes/luminosities derived from NB excess photometry:
\begin{align}
    &F_{\rm NB} = \Delta_{\rm NB}  \frac{f_{\rm NB} - f_{\rm BB}}{1 - (\Delta_{\rm NB} / \Delta_{\rm BB})},~{\rm and}\\
    &L_{\rm NB} = 4\pi d_L^2  F_{\rm NB},\nonumber
\end{align}
\noindent where $f_X$ is the flux density in erg s$^{-1}$ cm$^{-2}$ \AA$^{-1}$ for the relevant band, $\Delta$ is the FWHM of the wave band, and $d_L$ is the luminosity distance. For $\Delta$, the NB filter widths are reported in Table \ref{tab:filt_props}, and the broad-band filter widths are $\Delta R_{\rm C}$ = 1110 \AA\ for NB704/NB711, $\Delta i^{\prime}$ = 1419 \AA\ for NB816, and $\Delta z^{\prime}$ = 955 \AA\ for NB921/NB973. For $d_L$, we either use the spectroscopic redshift (where available) or the central redshift of the NB filter: $z=0.074$, 0.086, 0.242, 0.401, and 0.486 for NB704, NB711, NB816, NB921, and NB973, respectively. For dual NB704+NB711 \Ha\ emitters, we use the NB704 measurements since spectroscopy suggests that \Ha\ is more accurately captured by the NB704 filter than by the NB711 filter. These NB measurements require three types of corrections to determine \Ha\ SFRs: (1) NB filter profile, (2) \NII\ contamination, and (3) dust attenuation.

\subsubsection{NB filter profile correction}
\label{sec:NB}
The NB filter profiles are not flat, and thus the NB excess fluxes or luminosities will depend on the location of the \Ha\ emission line within
the filter:
\begin{equation}
  L_{\NII + {\rm H}\alpha} = f_{\rm filt}  L_{\rm NB}.
\end{equation}

\noindent
To correct for this effect, we use spectroscopic redshifts to determine the necessary correction relative to the centre of the filter. For excess emitters with spectroscopic redshift, the average (median) NB filter corrections $f_{\rm filt}$ for NB704, NB711, NB816, NB921, and NB973 emitters are 1.08 (1.03), 1.28 (1.07), 1.49 (1.20), 1.13 (1.02), and 1.33 (1.09), respectively. For galaxies without spectroscopic redshifts, we use a fixed statistical correction of 1.29, 1.41, 1.29, 1.33, and 1.30, respectively. These statistical corrections are derived from a weighted average based on the filter profiles.

The discrepancies between the average (median) corrections and the fixed statistical corrections likely have a number of possible reasons. First, many galaxies with spectroscopic follow-up tend to fall closer to the centre of the NB filter profile.
Second, selection bias effects may have skewed the redshift distribution towards a narrower one, as NB704 and NB921 dual emitters were targeted more often (higher spectroscopic completeness for two different samples). 
Third, cosmic variance may have also skewed our distribution, as spectroscopic follow-up (especially in NB921 and NB973 emitters) occurred within a small field of view. 
Finally, for some cases (e.g., the sample of NB704 or NB9711 emitters), we do not have many spectra, so outliers may heavily influence the redshift distribution of the spectroscopic sample.

\subsubsection{\NII\ correction}
\label{sec:NII}

%%
%% Figure 1
%%
\begin{figure}
	\centering
	\includegraphics[width=1\textwidth]{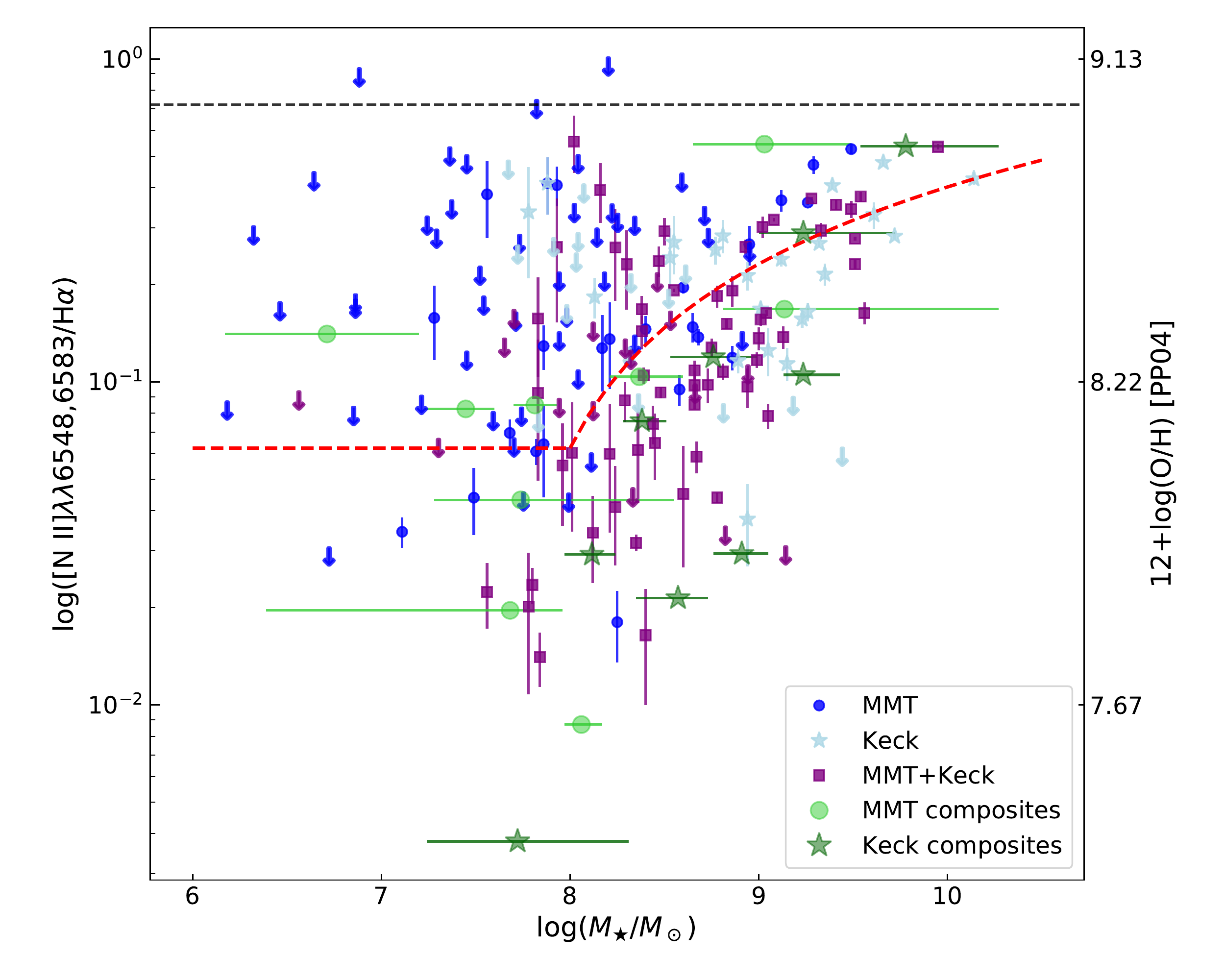}
	\vspace{-.5cm}
	\caption{The flux ratio of \NIIl\ to \Ha\ as a function of stellar mass. Since \NII$\lambda$6548 is often undetected, we adopt \NII$\lambda6583$/$\lambda6548$ = 2.96 \citep{Osterbrock06}. Arrows indicate upper flux limits for individual galaxies with S/N $<2$ on \NII$\lambda6583$. Horizontal bars indicate the mass range of the composite stacks. The dashed red line indicates the piecewise fit between the \NII/\Ha\ flux ratio and $\Mstar$ (Equation~\ref{eqn:nii1}). The dashed black line indicates the maximum value of \NII$\lambda6583$/\Ha\ = 0.54 above which galaxies are likely AGNs instead of star-forming galaxies \citep{Kennicutt08}.}
	\label{fig:nii_ha}
\end{figure}

The NB filters are wide enough ($\sim$100 \AA) to include \NIIl\ emission (hereafter \NII). To correct for this contribution, we use emission-line measurements from individual galaxies (S/N $\geq2$ on \NII$\lambda6583$) and average measurements from composite spectra (see Section~\ref{sec:stack}). The composite spectra used to apply these corrections consist of measurements from Keck $\Mstar$--$z$ stacks, and measurements from MMT $\Mstar$--$z$ stacks where Keck measurements are not available (e.g., sources detected in just the NB704, NB711, NB816 filters). We illustrate these \NII\ measurements in Figure~\ref{fig:nii_ha}. For the composite spectra, a least-squares fit was determined between the \NII/\Ha\ flux ratio and $\Mstar$. This least-squares fit, also illustrated in Figure~\ref{fig:nii_ha}, is described as:
\begin{align}
  \label{eqn:nii1}
  &\frac{\NII}{\Hyd\alpha} = 0.062~(x \leq 8.0),~{\rm and}\\
  &\frac{\NII}{\Hyd\alpha} = 0.169 x - 1.293 ~(x > 8.0),\nonumber
\end{align}
where $x \equiv \log(\Mstar/\Msun)$. Since \NII$\lambda$6548 is weak, we assume a flux ratio of $\lambda$6583/$\lambda$6548 = 2.96 \citep{Osterbrock06}. 
Thus, the observed \Ha\ luminosity is:
\begin{align}
  L_{{\rm H}\alpha} = \frac{1}{1+\NII/\Hyd\alpha}  L_{\NII + {\rm H}\alpha}.
\end{align}

\subsubsection{Dust attenuation correction from Balmer decrements}
\label{sec:dust}

%%
%% Figure 2
%%
\begin{figure*}
	\centering
	\includegraphics[width=1\textwidth]{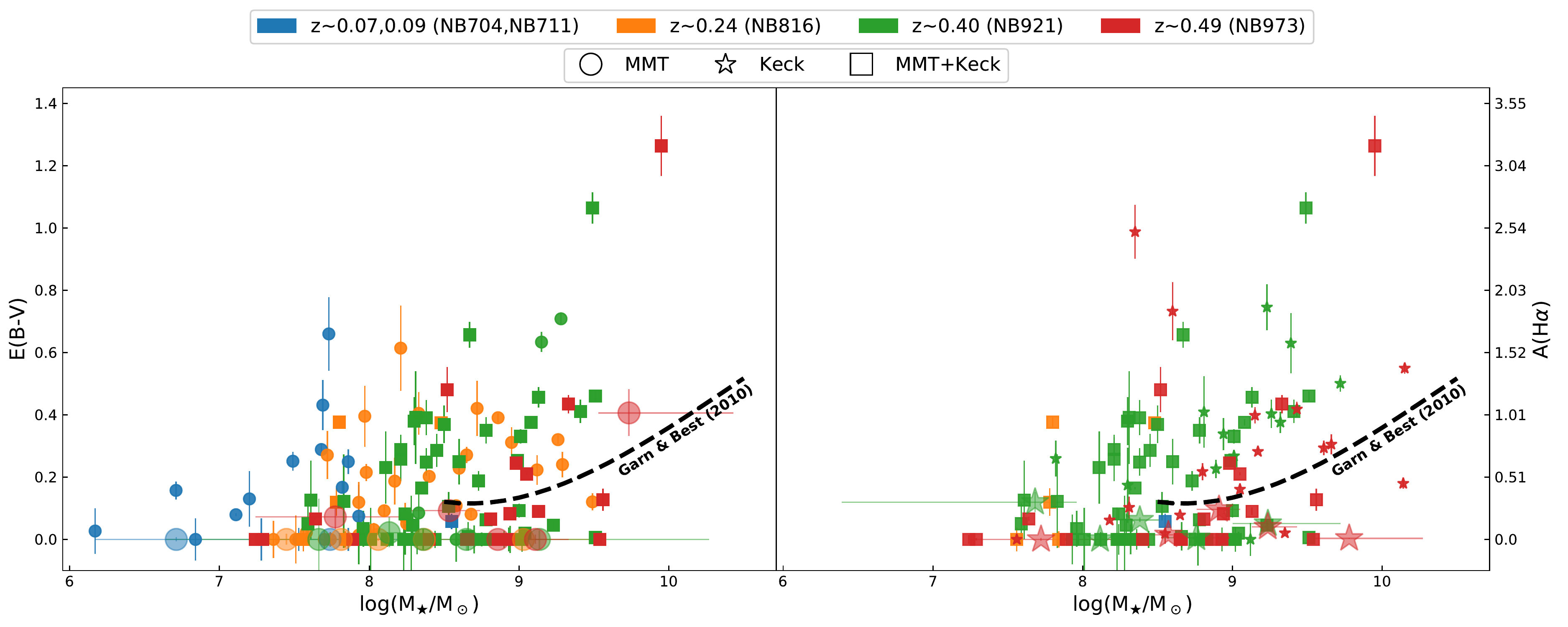}
	\vspace{-.5cm}
	\caption{Dust attenuation color excess derived from Balmer decrements for individual sources (small filled shapes) and composite $\Mstar$--$z$ stacks (large transparent shapes). The \Ha/\Hb\ Balmer decrements are plotted where they exist; otherwise the \Hg/\Hb\ Balmer decrements are plotted. Measurements from MMT, Keck, and a combination of both are indicated as circles, stars and squares, respectively. The relationship between \EBVm\ and stellar mass for local galaxies \citep{Garn10} is illustrated by the black dashed line.}
	\label{fig:mstar_vs_ebv}
\end{figure*}

The ``gold standard'' of dust attenuation correction is the use of Balmer decrements, the \Ha/\Hb\ and H$\gamma$/\Hb\ flux ratios from spectroscopic measurements \citep{Osterbrock06,Kennicutt12}.
The advantage here is that the Balmer decrements trace the dust associated with the source of \Ha\ emission, the ionized gas. Our sample has deep spectroscopy for $\approx$34\% of the \Ha\ emitting sample, and reliable individual Balmer line measurements for $\approx$11\% of the sample. Through stacked measurements of emission lines from galaxies with spectroscopic coverage (Section~\ref{sec:stack}), we are able to use Balmer decrements to determine dust attenuation corrections for the entire sample.
Here, the color excess $E$($B$--$V$) for nebular gas and amount of attenuation at \Ha\ are given by
\begin{align}
  & \log\left[\frac{(\Hyd n /\Hbe)_{\rm obs}}{(\Hyd n /\Hbe)_0}\right] = -0.4\EBV[k(\Hyd n)-k(\Hbe)],{\rm ~and}\\
  & A({\rm H}\alpha) \equiv k({\rm H}\alpha)  \EBV,\nonumber
\end{align}
where H$n$ is either \Ha\ or H$\gamma$, $k$(H$n$) is the attenuation coefficient following
\citet{Cardelli89}\footnote{$k$(\Ha) = 2.535, $k$(\Hb) = 3.609, and $k$(H$\gamma$) = 4.173.}, and the intrinsic Balmer decrements are
(\Ha/\Hb)$_0=2.86$ and (H$\gamma$/\Hb)$_0=0.468$ assuming Case B recombination with an electron temperature of 10$^4$ K.

We illustrate $\EBV$ and $A({\rm H}\alpha)$ for individual galaxies and for composite stacks in Figure~\ref{fig:mstar_vs_ebv}. Here,
Balmer decrement values from the composite $\Mstar$--$z$ spectra are used as they most accurately describe what the average dust attenuation
is as a function of stellar mass and redshift.
The majority of the individual sources have significantly higher \EBVm\ values than predicted by the empirical \EBVm--$\Mstar$ relation for local galaxies \citep{Garn10}, indicating that adopting this local relation is not ideal for our sample.

For galaxies without spectra or with less reliable detections of the Balmer lines, we use the composite measurements as follows.
    For NB704, NB711 and NB816 \Ha\ emitters, we use the corresponding \Ha/\Hb\ Balmer decrement from the MMT $\Mstar$--$z$ composites.
    For NB921 \Ha\ emitters, we use the Keck (MMT) \Ha/\Hb\ Balmer decrements for stellar masses of $\Mstar \lesssim 5\times10^{9}$
    ($\Mstar \gtrsim 5\times10^9$) $\Msun$. Finally for NB973 \Ha\ emitters, we use the Keck \Ha/\Hb\ Balmer decrements.
After applying the dust attenuation corrections, the dust-corrected \Ha\ luminosity is:
\begin{equation}
  \log\left(L_{{\rm H}\alpha,{\rm corr}}\right) = \log\left(L_{{\rm H}\alpha}\right) + 0.4 A({\rm H}\alpha).
\end{equation}

\subsubsection{Deriving SFRs}
\label{sec:ha_sfrs}
\noindent

Assuming a \citet{Chabrier03} initial mass function (IMF) with minimum and maximum masses of 0.1 and 100 $\Msun$ and solar metallicity \citep{Kennicutt98}, the SFR can be estimated from the dust-corrected \Ha\ luminosity as
\begin{equation}
  \frac{{\rm SFR}}{\Msun\ {\rm yr}^{-1}} = 4.4\times10^{-42} \cdot \frac{L_{{\rm H}\alpha,\rm corr}}{{\rm erg}~{\rm s}^{-1}}.
\end{equation}
However, the majority of our galaxies are at low stellar masses, which suggests that they have low metal abundances. Following \cite{Ly16}, we use a metallicity-dependent SFR--$L$(\Ha) conversion:
\begin{equation}
  \log\left[\frac{{\rm SFR}(\Msun\ {\rm yr}^{-1})}{L_{{\rm H}\alpha,\rm corr}({\rm erg}~{\rm s}^{-1})}\right] = -41.34 + 0.39y + 0.127y^2,
\end{equation}
where $y = \log({\rm O/H}) + 3.31.$\footnote{$y = 0$ for solar metallicity, $12+\log({\rm O/H}) = 8.69$}
This relation was determined using \Ha\ luminosity predictions from Starburst99 spectral synthesis models \citep{Leitherer99}. Here, we adopt the Padova stellar tracks \citep{Bressan93,Fagotto94}, a constant star formation history, and metallicities of 0.02, 0.20, 0.40, 1.0, and 2.5 $Z_{\sun}$. The original estimates were based on a \cite{Kroupa01} IMF, and were transformed to a \cite{Chabrier03} IMF with an offset of 0.1 dex.
This relation is valid between 0.02$Z_{\sun}$ and 2.5$Z_{\sun}$ or $12+\log({\rm O/H})=7.0$ and 9.1, and thus is applicable to our entire sample. To estimate the oxygen abundances for our galaxies, we use the \NII$\lambda$6583/\Ha\ flux ratio either from individual or composite measurements (see Figure~\ref{fig:nii_ha}). Specifically, we use the \cite{PP04} cubic calibration.

After correcting for the NB filter profile, \NII\ contamination, and dust attenuation, the \NHa\ identified \Ha\ emitting NB excess galaxies have SFRs ranging between $10^{-3.57} \Msun\ {\rm yr}^{-1}$ to $10^{1.40} \Msun\ {\rm yr}^{-1}$ (mean: $10^{-1.46} \Msun\ {\rm yr}^{-1}$).

\subsection{Stellar masses from SED modeling}
\label{sec:SED}
\noindent
The Fitting and Assessment of Synthetic Templates (FAST; \citealp{Kriek09}) code was used to determine physical properties of the full NB excess emitter sample by modeling their optical-to-infrared SEDs. FAST fits the broadband photometric data points with modeled fluxes from stellar synthesis models. In our SED modeling, we use either spectroscopic redshifts or central redshifts for the relevant NB filter (see Table~\ref{tab:filt_props}).
Photometric data are obtained from running SExtractor \citep[version 2.5.0;][]{Bertin96} with NB images for source detection. These galaxies have also been imaged in the \textit{FUV} and \textit{NUV} bands with \textit{GALEX} \citep{Martin05}; for these measurements, more accurate photometry is obtained by point spread function fitting with \textsc{iraf/daophot} \citep[version 2.16;][]{Stetson87}. Further \textit{GALEX} imaging and photometric measurement details are available in \citet{Ly09}, \citet{Ly11}, and \citet{Ly16}.

%%
%% Figure 3
%%
\begin{figure}
	\centering
	\includegraphics[width=1\textwidth]{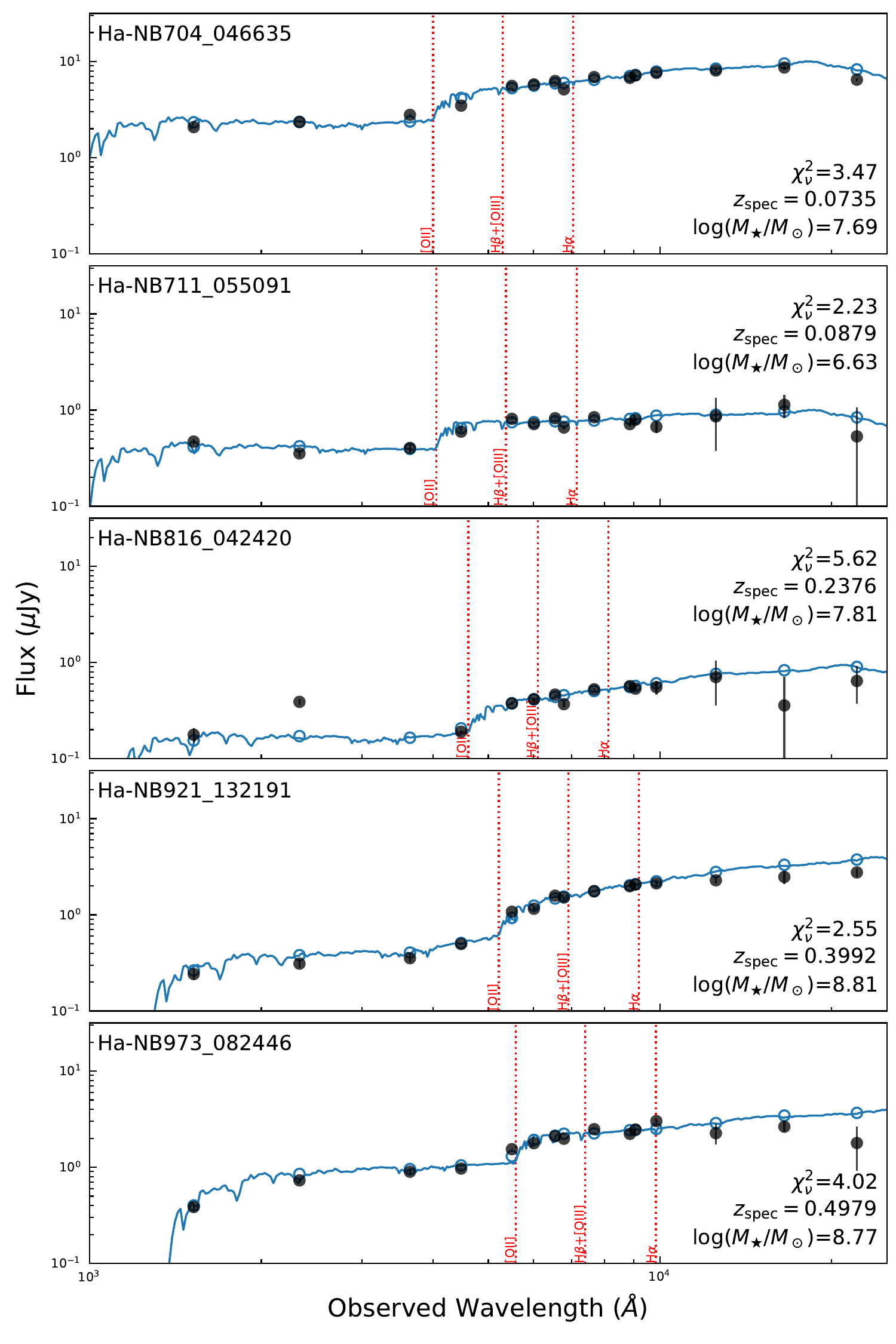}
	\vspace{-0.75cm}
	\caption{SEDs (grey filled circles) and the best-fitting results from modeling the SED (solid blue line and open blue circles) in each of the five NB filters. These SEDs were randomly selected. 
		Vertical red dashed lines illustrate the location of common spectral features. Good agreement is seen between photometric and modeled fluxes.}
	\label{fig:sed_fits}
\end{figure}

In FAST, \citet{Bruzual03} stellar synthesis models, with an exponentially declining star formation history, SFR~$\propto$~$\exp(-t/ \tau)$, were used. We chose $\log(\tau/{\rm yr})$ = 7.0, 8.0, 9.0, and 10.0 in order to include bursty, intermediate, and roughly constant star formation histories. The \citet{Calzetti00} dust attenuation relation was adopted, and we assumed a \citet{Chabrier03} initial mass function (IMF) and a stellar metallicity of $Z = 0.02$.

The grid of models FAST considered also included dust reddening ($A_V$) at 0.0--3.0 mag in 0.1 mag increments, stellar ages of $\log({\rm age/yr})=7.0$--10.1 in 0.1 dex increments, and redshifts from 0.01--1.70 in 0.001 increments. Some of the physical properties that were determined include best-fitting stellar ages, $A_V$, SFR, stellar masses, and $\tau$ values, as well as their respective confidence intervals. These estimates were determined from $\chi^2$ minimization; FAST determines every point of the model grid.

A sample of SED fits, for each NB sample, can be seen in Figure~\ref{fig:sed_fits}. The SED models fit the data points well within error bars. Identified \Ha\ emitting NB excess galaxies have stellar masses between $10^{4.3}$ to $10^{10.4}~\Msun$ (mean: $10^{7.96}~\Msun$). Galaxies corresponding to the low end of these stellar masses were ultimately not included in our analysis (Section~\ref{sec:results} for discussion).

\section{Results \& Discussion}
\label{sec:results}
\noindent

We discuss four key results from our study: comparison between UV and \Ha\ SFRs (Section~\ref{sec:uv_sfrs}), the \Ha\ specific SFR and its evolution with redshift and dependence with stellar mass (Section~\ref{sec:sSFR}), the SFR--$\Mstar$ relation (Section~\ref{sec:mainseq}), and the dispersion of the SFR--$\Mstar$ relation (Section~\ref{sec:dispersion}). We also examine the completeness of our survey in Section~\ref{sec:comp}.

The following results and analyses required that we further restrict our sample for the following reasons. First, due to a large amount of galaxies detected near the 3$\sigma$ flux limit, a higher sigma selection (4$\sigma$) was imposed on NB excess fluxes. The new selection reduces the sample to 730 galaxies. The revised 4$\sigma$ NB excess flux selections correspond to $7.3\times10^{-18}$ (NB704), $8.4\times10^{-18}$ (NB711), $5.9\times10^{-18}$ (NB816), $7.8\times10^{-18}$ (NB921), and ${2.3\times10^{-16}}$~erg~s$^{-1}$~cm$^{-2}$ (NB973).

Additionally, we impose a low-mass cutoff of $10^6 \Msun$ as none of the galaxies below this cutoff were spectroscopically confirmed, and many of them had unreliable photometric redshifts from \textsc{eazy}. In fact, a number of these low-mass galaxies had photometric redshifts suggestive of different emission lines corresponding to higher redshifts and hence higher stellar masses. This new selection, with both the 4$\sigma$ cut and the low-mass limit cutoff of $10^6 \Msun$, reduces the sample to \NHaF\ galaxies.

In our analyses, we randomise each data point in our observations 10,000 times, assuming a Gaussian with the standard deviation equal to the measurement uncertainty. Each randomization is re-fit with \texttt{scipy.optimize.curve\_fit}.
Uncertainties for reported averages are derived using the bootstrap approach, where we randomly select galaxies and compute the average 10,000 times.
Randomization accounts for measurement uncertainties, examines degeneracies in our fitting parameters, and minimizes the number of underlying assumptions that are adopted.

\vspace{\baselineskip}
\subsection{Comparisons between \Ha\ and UV SFRs}
\label{sec:uv_sfrs}

%%
%% Figure 4
%%
\begin{figure*}
  \centering
  \includegraphics[width=1\textwidth]{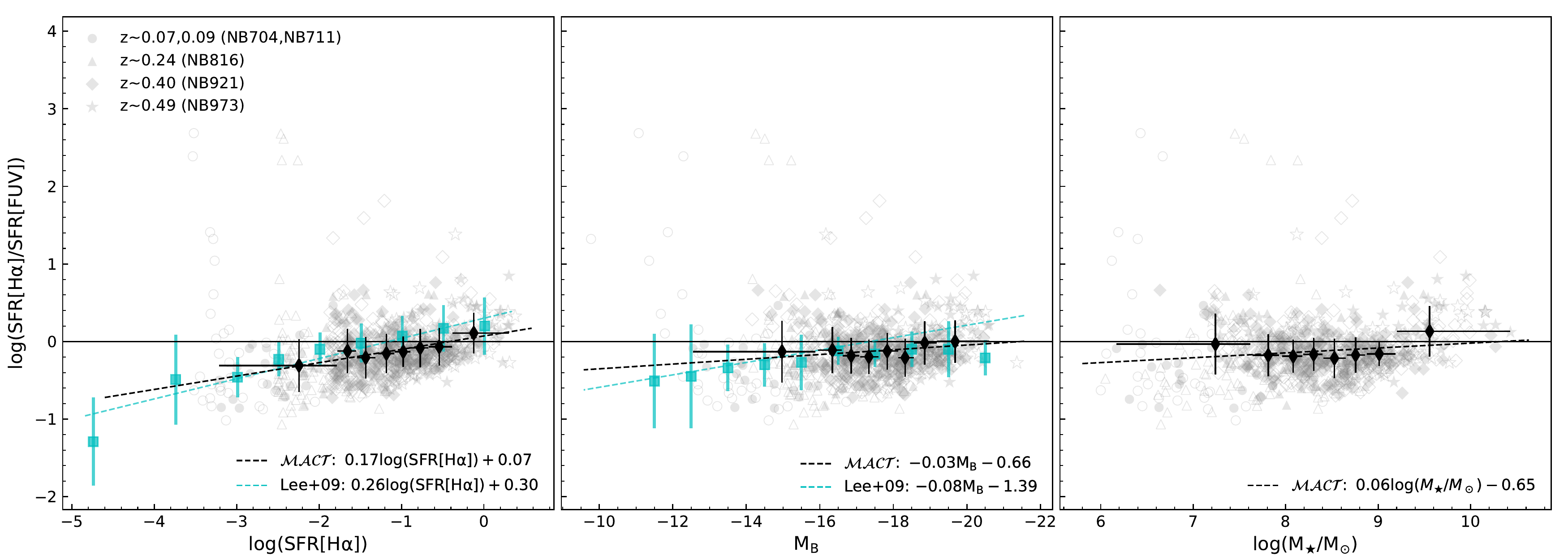}
  \caption{Ratio of {\it observed} \Ha-to-FUV SFR against the \Ha\ SFR (left panel), \textit{B}-band absolute magnitude (middle panel), and stellar mass (right panel). Empty (filled) symbols represent galaxies without (with) spectroscopic confirmation. Average values in each 12.5-percentile bin of galaxies with spectroscopic confirmation are represented with solid black diamonds, with 1$\sigma$ scatter plotted as well. A linear least squares fit to the data is overlaid with a dashed black line, with the best-fitting line for each dataset given in the lower right of each panel. For the first two panels, data and results from \citet{Lee09} are plotted in cyan. A \citet{Chabrier03} IMF is adopted here.}
  \label{fig:sfr_ratio}
\end{figure*}

In addition to \Ha-derived SFRs, our photometric data provide FUV continuum measurements, allowing us to compare two independent SFR estimates.
Under the assumption of a continuous star formation history, these two SFR measurements, which probe different time-scale ($\lesssim$5 Myr for \Ha\ and $\lesssim$100 Myr for FUV), should agree. However, for low-SFR galaxies, a systematic discrepancy between \Ha- and FUV-derived SFRs has been observed, wherein the \Ha\ indicator underpredicts the total SFR compared to the FUV \citep[e.g.,][]{Lee09}. One explanation of this discrepancy is that, at low SFRs, statistically sampling the IMF will not produce enough massive OB stars, resulting in a deficit for \Ha\ measurements \citep{Weidner13}.

To derive UV SFRs, we follow Section~\ref{sec:ha_sfrs} where we use the following metallicity-dependent $L$(FUV)--SFR relation derived from Starburst99 estimates:
\begin{equation}
\log\left[\frac{{\rm SFR}(\Msun\ {\rm yr}^{-1})}{L_{\nu}(1500)({\rm erg}~{\rm s}^{-1}~{\rm Hz}^{-1})}\right] = -28.179 + 0.151y + 0.053 y^2.
\end{equation}
Here we also adopt a \cite{Chabrier03} IMF with minimum and maximum masses of 0.1 and 100 $\Msun$.
The 1500\AA\ luminosities were obtained from interpolating our best-fitting SEDs derived from our FAST modeling (see Section~\ref{sec:SED}).

\begin{figure}
	\centering
	\includegraphics[width=0.95\textwidth]{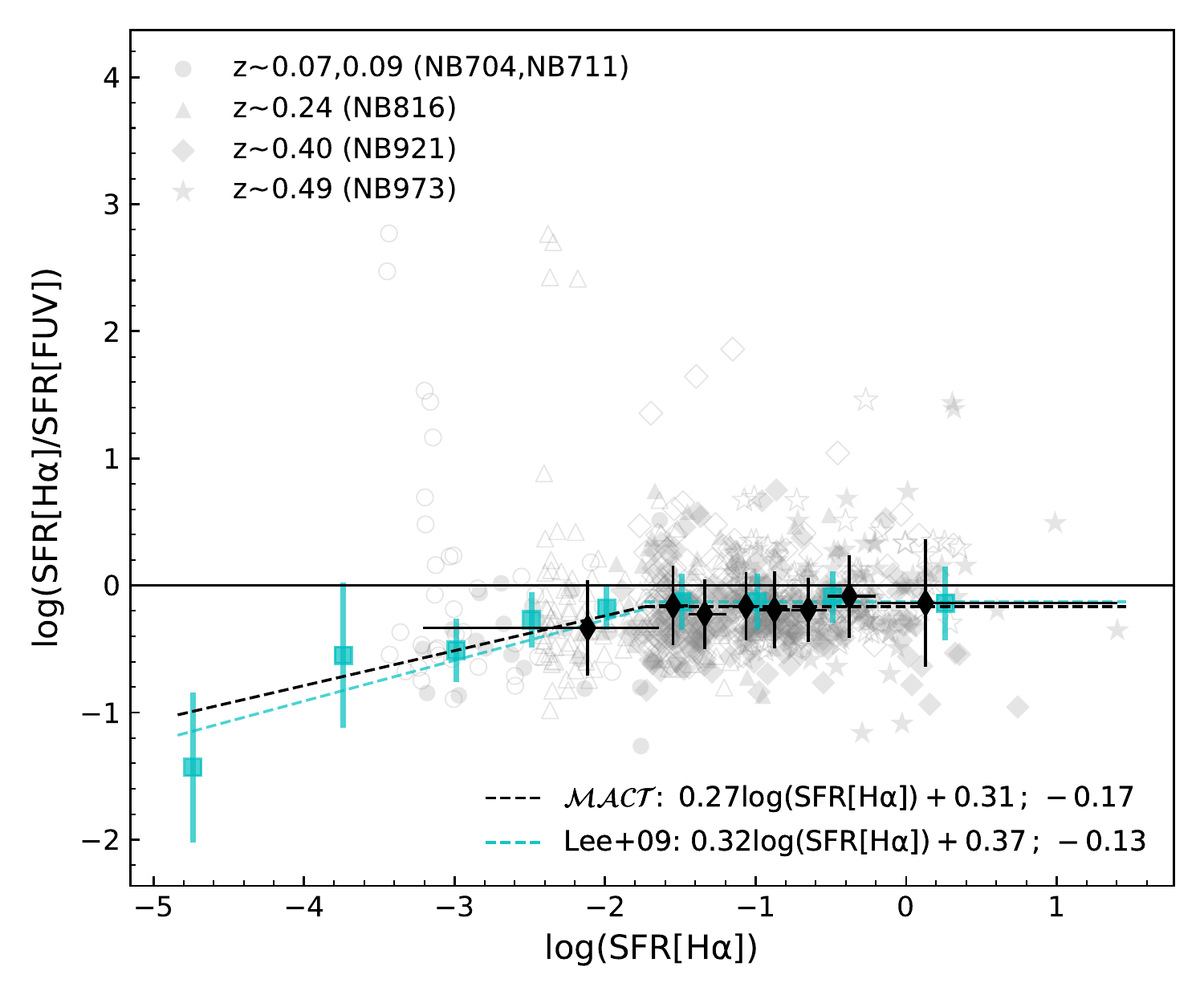}
	\caption{Similar to the left panel of Figure~\ref{fig:sfr_ratio}, but for {\it dust-corrected} SFRs.
		A piecewise fit to the data is overlaid with a black dashed line and labelled in the lower right corner.
		The form of the fit is a linear relation for log(SFR[\Ha]) $ \leq -1.74$, and a constant at higher SFR.
		Good agreement is seen with respect to \citet{Lee09}.}
	\label{fig:sfr_ratio_dustcorr}
\end{figure}

Under the assumption that the FUV luminosity is a more robust indicator of star formation in low-mass galaxies than \Ha, we can systematically correct our \Ha\ SFRs.
For robustness, we only consider galaxies with spectroscopic confirmation.
As demonstrated in Figure~\ref{fig:sfr_ratio}, the {\it observed} \Ha-to-FUV SFR ratio shows a deficit against the \Ha\ SFR, \textit{B}-band absolute magnitude, and stellar mass.
This deficit can be described as follows:
\begin{equation}
  \log\left(\frac{{\rm SFR[H\alpha]}}{{\rm SFR[FUV]}}\right) = 0.17\psi + 0.07,
  \label{eqn:fuv_corr}
\end{equation}
where $\psi \equiv \log({\rm SFR[H\alpha]})$. Correcting our SFRs for dust attenuation,\footnote{Here we use the nebular reddening from the Balmer decrement, ${\EBV_{\rm gas}}$, adopt a scaling factor of 2.27 between nebular and stellar reddening, and assume \citet{Calzetti00} reddening for the FUV measurements.} 
which is illustrated in Figure~\ref{fig:sfr_ratio_dustcorr}, we find good agreement with \citet{Lee09}.
Our values of observed and dust-corrected \Ha-to-FUV SFR ratios against the \Ha\ SFRs, per 12.5-percentile bin of galaxies, are tabulated in Table~\ref{tab:uv_ha_sfr}.

%%
%% Figure 5
%%

\begin{table}[ht]
  \centering
  \caption{Summary of UV--\Ha\ SFR comparisons}
  \label{tab:uv_ha_sfr}
  \begin{threeparttable}
    \centering
    \begin{tabular}{lcc}
      \hline\hline
      $\log{\left({\rm SFR[H\alpha]}\right)}$ & $N_{\rm spec}$ &
      $\log\left(\frac{{\rm SFR[H\alpha]}}{{\rm SFR[FUV]}}\right)$ \\
      (1) & (2) & (3)\\ \hline
      \multicolumn{3}{c}{Observed}\\

--2.24$_{-0.97}^{+0.46}$ & 45 & --0.31$_{-0.55}^{+0.89}$ \\
--1.65$_{-0.12}^{+0.10}$ & 44 & --0.12$_{-0.50}^{+0.72}$ \\
--1.43$_{-0.11}^{+0.12}$ & 44 & --0.21$_{-0.41}^{+0.87}$ \\
--1.18$_{-0.10}^{+0.10}$ & 45 & --0.15$_{-0.53}^{+0.54}$ \\
--0.98$_{-0.10}^{+0.09}$ & 44 & --0.13$_{-0.44}^{+0.44}$ \\
--0.77$_{-0.11}^{+0.10}$ & 44 & --0.08$_{-0.41}^{+0.70}$ \\
--0.54$_{-0.13}^{+0.15}$ & 44 & --0.07$_{-0.46}^{+0.83}$ \\
--0.12$_{-0.26}^{+0.43}$ & 45 & +0.11$_{-0.38}^{+0.74}$ \\

	  \noalign{\vskip 1mm}
	  \hline
      \multicolumn{3}{c}{Dust-corrected}\\
      
--2.12$_{-1.09}^{+0.47}$ & 45 & --0.33$_{-0.93}^{+1.08}$ \\
--1.55$_{-0.09}^{+0.10}$ & 44 & --0.16$_{-0.51}^{+0.74}$ \\
--1.34$_{-0.11}^{+0.15}$ & 44 & --0.23$_{-0.46}^{+0.80}$ \\
--1.06$_{-0.12}^{+0.09}$ & 45 & --0.16$_{-0.67}^{+0.50}$ \\
--0.88$_{-0.10}^{+0.10}$ & 44 & --0.19$_{-0.67}^{+0.94}$ \\
--0.65$_{-0.11}^{+0.13}$ & 44 & --0.19$_{-0.57}^{+0.70}$ \\
--0.38$_{-0.14}^{+0.18}$ & 44 & --0.09$_{-1.07}^{+0.77}$ \\
+0.13$_{-0.32}^{+1.27}$ & 45 & --0.14$_{-0.95}^{+1.57}$ \\

	  \hline
    \end{tabular}
    \begin{tablenotes}
      \item \leavevmode\kern-\scriptspace\kern-\labelsep (1): The mean value of the logarithm of the \Ha\ SFR values per 12.5-percentile bin of galaxies. The range of the bin is denoted as errors.  (2): Number of spectroscopic measurements. (3): The mean value of the logarithm of ratio of \Ha-to-FUV SFRs. The 1$\sigma$ scatter is denoted as errors.
      Observed values are illustrated in the left-hand panel of Figure~\ref{fig:sfr_ratio}. Dust-corrected values are illustrated in Figure~\ref{fig:sfr_ratio_dustcorr}, with dust attenuation corrections based on Balmer decrements (see Section~\ref{sec:dust}).
    \end{tablenotes}
  \end{threeparttable}
\end{table}

Our sample and the sample in \citet{Lee09} were independently selected using different methods at different redshifts; the agreement between these studies supports the prediction in \citet{Lee09} whereby the semi-empirical integrated galactic IMF model \citep{Kroupa03} could account for the systemic underprediction of \Ha-derived SFRs \citep{Weidner13}.
In the work of \citet{Lee09}, they fit a piece-wise function with turnover at
--1.5.\footnote{For a \cite{Salpeter55} IMF. It is --1.74 for a \cite{Chabrier03} IMF.}
Following their approach, our resulting fit is:
\begin{align}
&\log\left(\frac{{\rm SFR[H\alpha]}}{{\rm SFR[FUV]}}\right) = 0.27\psi + 0.31~(\psi \leq -1.74),~{\rm and}\\
&\log\left(\frac{{\rm SFR[H\alpha]}}{{\rm SFR[FUV]}}\right) = -0.17~(\psi > -1.74).\nonumber
\end{align}
To correct for this underprediction, we use the best-fitting relation between the dust-corrected \Ha-to-FUV
SFR ratio and the \Ha\ SFR. The resulting ``true'' \Ha\ SFR is then:
\begin{equation}
\psi_{\rm true} = \psi - \log\left(\frac{{\rm SFR[H\alpha]}}{{\rm SFR[FUV]}}\right).
\end{equation}
In our analyses below, we report results with the \Ha\ underprediction correction.

\vspace{\baselineskip}
\subsection{\Ha\ Specific SFRs}
\label{sec:sSFR}
\noindent
The SFR per unit stellar mass, or specific SFR (sSFR), is often used as an indicator of active star formation as it is related to the gas depletion time-scale \citep[e.g.,][]{Kennicutt98,Kennicutt12}. Studies have examined how sSFR evolves toward higher redshift below $z \lesssim 2$ \citep[e.g.,][]{Lehnert15,Tasca15} and how it depends on stellar mass \citep[e.g.,][]{Salim07,Reyes15,Davies19}. However, these studies mostly probe higher mass galaxies, and have not examined evolution of galaxies below stellar masses of 10$^9 \Msun$.

To address this lack of understanding, we first use the $\mathcal{MACT}$ sample to determine the evolution of sSFR at a given stellar mass, $\log(\Mstar/\Msun) = 8.0 \pm 0.5$. We selected this stellar mass for two reasons: (1) it is near the average stellar mass of the reduced $\mathcal{MACT}$ sample, $10^{8.3}~\Msun$, and (2) all NB redshift bins, from $z=0.07$ to 0.5, are included in this analysis. We explore two redshift dependencies and find significant evolution: $\log({\rm sSFR}) \propto A \log(1+z)$ with $A = 5.26\pm 0.75$, and $\log({\rm sSFR}) \propto B z$ with $B = 1.83\pm 0.26$. This result is consistent with the previously mentioned studies of higher mass galaxies that have found, at fixed stellar mass, increased sSFR evolution toward higher redshift over our redshift range.

%%
%% Figure 6
%%
\begin{figure}
	\centering
	\includegraphics[width=1\textwidth]{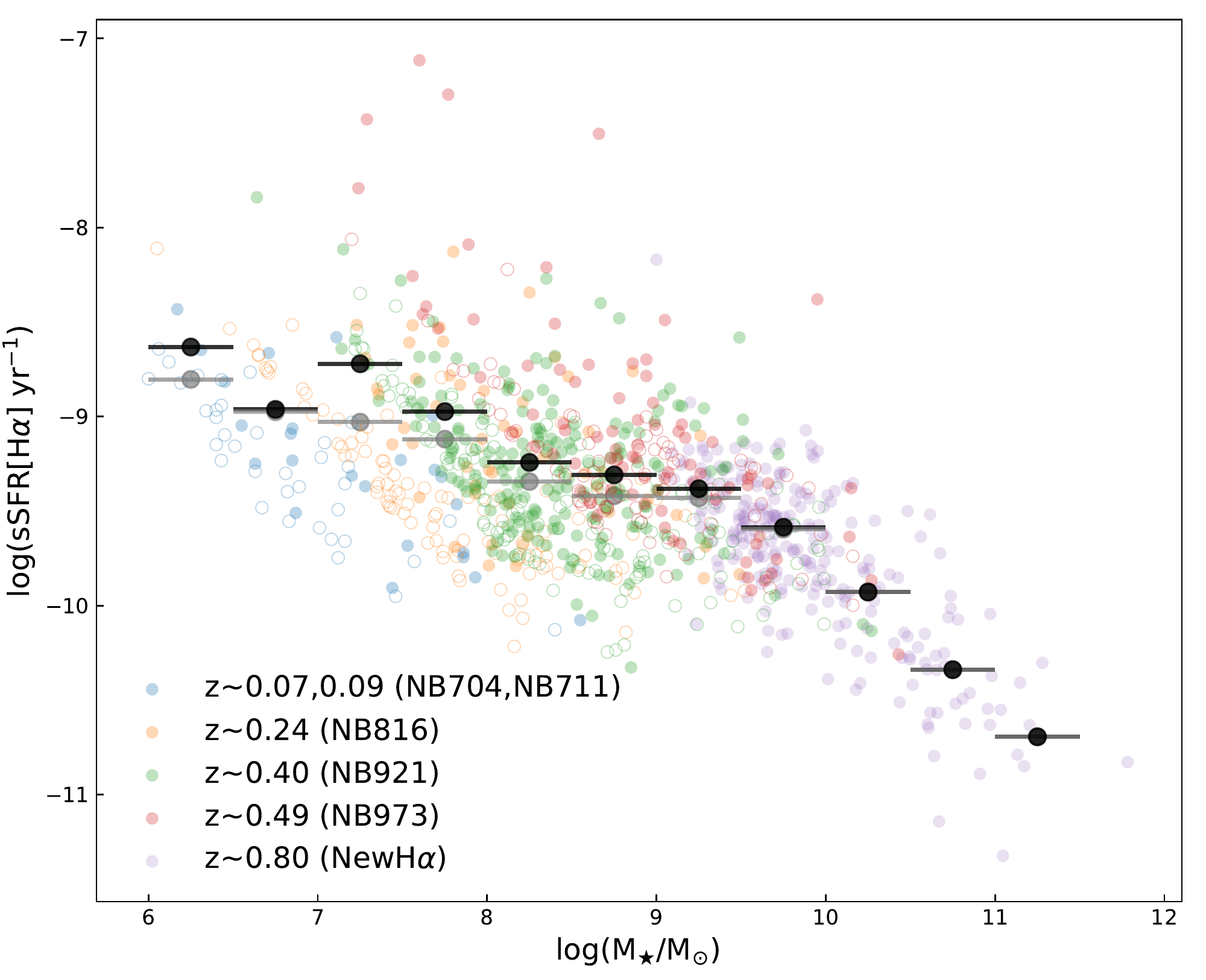}
	\vspace{-0.7cm}
	\caption{Specific SFRs of NB-selected \Ha\ galaxies as a function of stellar mass, spanning six dex in stellar mass and half the age of the Universe (up to $z\approx0.8$).
	Colored empty (filled) symbols represent galaxies without (with) spectroscopic confirmation.
	Gray (black) circles represent the average sSFR for all sources (sources with spectroscopic confirmation) in each stellar mass bin.
	This figure includes measurements from the New\Ha\ survey \citep{Reyes15}. 
	}
	\label{fig:mainseq_sSFRs}
\end{figure}

%%
%% Figure 7
%%
\begin{figure*}
	\centering
	\includegraphics[width=0.99\textwidth]{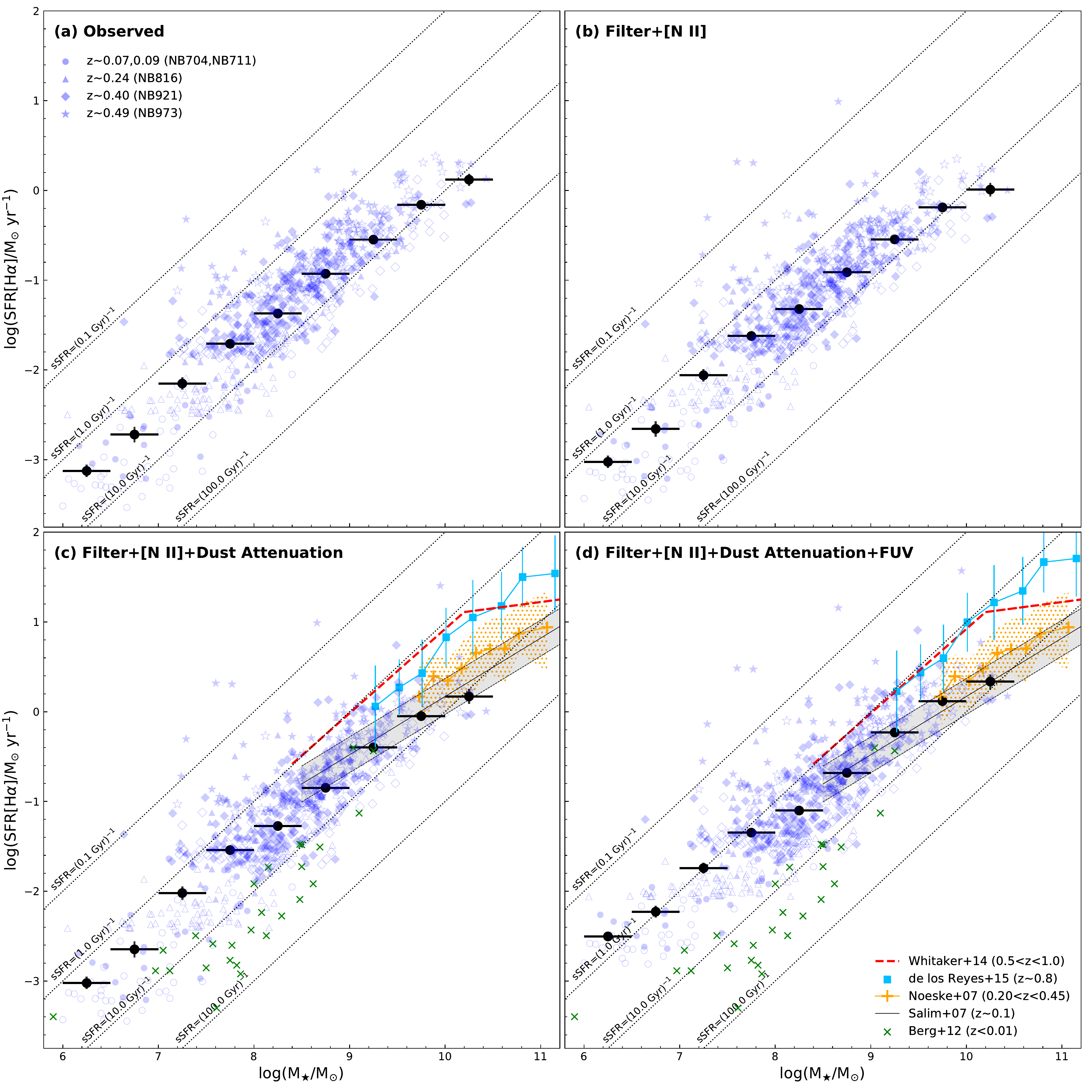}
	\vspace{-0.5cm}
	\caption{The SFR--$\Mstar$ relation for \Ha\ emitters between $z=0.07$ and $z=0.5$.
		Empty (filled) symbols represent galaxies without (with) spectroscopic confirmation. (a) the observed sample without any applied corrections, and (b), (c), and (d) have different corrections applied, as indicated in the figure. Average SFRs in each stellar mass bin are represented with solid black dots. Results from other empirical studies are overlaid as follows: \citet{Whitaker14} with dashed red lines, NewH$\alpha$ measurements from \citet{Reyes15} with cyan squares, \citet{Noeske07} with orange pluses, \citet{Salim07} with a solid black line, and \citet{Berg12} with green crosses.
	}
	\label{fig:mainseq}
\end{figure*}

Next, we examine the correlation of sSFR with stellar mass at $\lesssim 3\times10^9 \Msun$, which is illustrated in Figure~\ref{fig:mainseq_sSFRs}. Since our study is unable to probe high stellar masses ($\gtrsim3\times10^9 \Msun$), we include measurements from the NewH$\alpha$ survey \citep{Ly11, Lee12}.
The NewH$\alpha$ survey identified 394 \Ha\ emitting galaxies using a near-infrared NB filter in several deep fields for a total of 0.82~deg$^2$ with stellar masses of $3\times10^9$--$10^{11}~\Msun$. It probes higher redshifts ($z\approx0.8$) down to an emission-line flux of ${2\times10^{-17}}$~erg~s$^{-1}$~cm$^{-2}$ (3$\sigma$) or an observed \Ha\ SFR of $0.3 \Msun$~yr$^{-1}$ at the low-luminosity end, and identified galaxies with a minimum \Ha\ rest-frame equivalent width (EW) of 22 \AA\ at the high-mass end.
These \Ha-selected galaxies with spectroscopic confirmation (N = 299) are illustrated in Figure~\ref{fig:mainseq_sSFRs} in purple. Here we re-derived \Ha\ SFRs for the New\Ha\ sample by using our metallicity-dependent SFR relation. To estimate metallicity, we used the tabulated \NII/\Ha\ ratio from \cite{Reyes15}.

With the combined $\mathcal{MACT}$ and NewH$\alpha$ datasets, we find: (1) low-mass galaxies have higher sSFRs than high-mass (${\gtrsim 3\times 10^9 \Msun}$) galaxies, and (2) the sSFRs appear to plateau for low-mass galaxies unlike high-mass galaxies where there is a strong negative dependence. While the sSFR--$\Mstar$ relation flattens toward lower stellar masses, it is not entirely a constant. Thus, this result suggests that the SFR--$\Mstar$ relation has a slope less than unity for most stellar masses. This result is consistent with many studies such as \cite{Salim07} and \cite{Davies19} for local Universe galaxies, \cite{McGaugh17} for low surface brightness local galaxies, and \cite{Boogaard18} for field star-forming galaxies (see Section~\ref{sec:other_studies}).
This trend is less evident for galaxies with spectroscopic confirmation, but we caution this may be a selection effect due to spectroscopic constraints of emission line detections. We note that these results are not due to an incompleteness of our study, as higher sSFR galaxies would have easily been detected with our narrowband photometry.

\subsection{Star Formation Sequence Relation}
\label{sec:mainseq}

%%
%% Figure 8
%%
\begin{figure}
  \centering
  \includegraphics[width=1\textwidth]{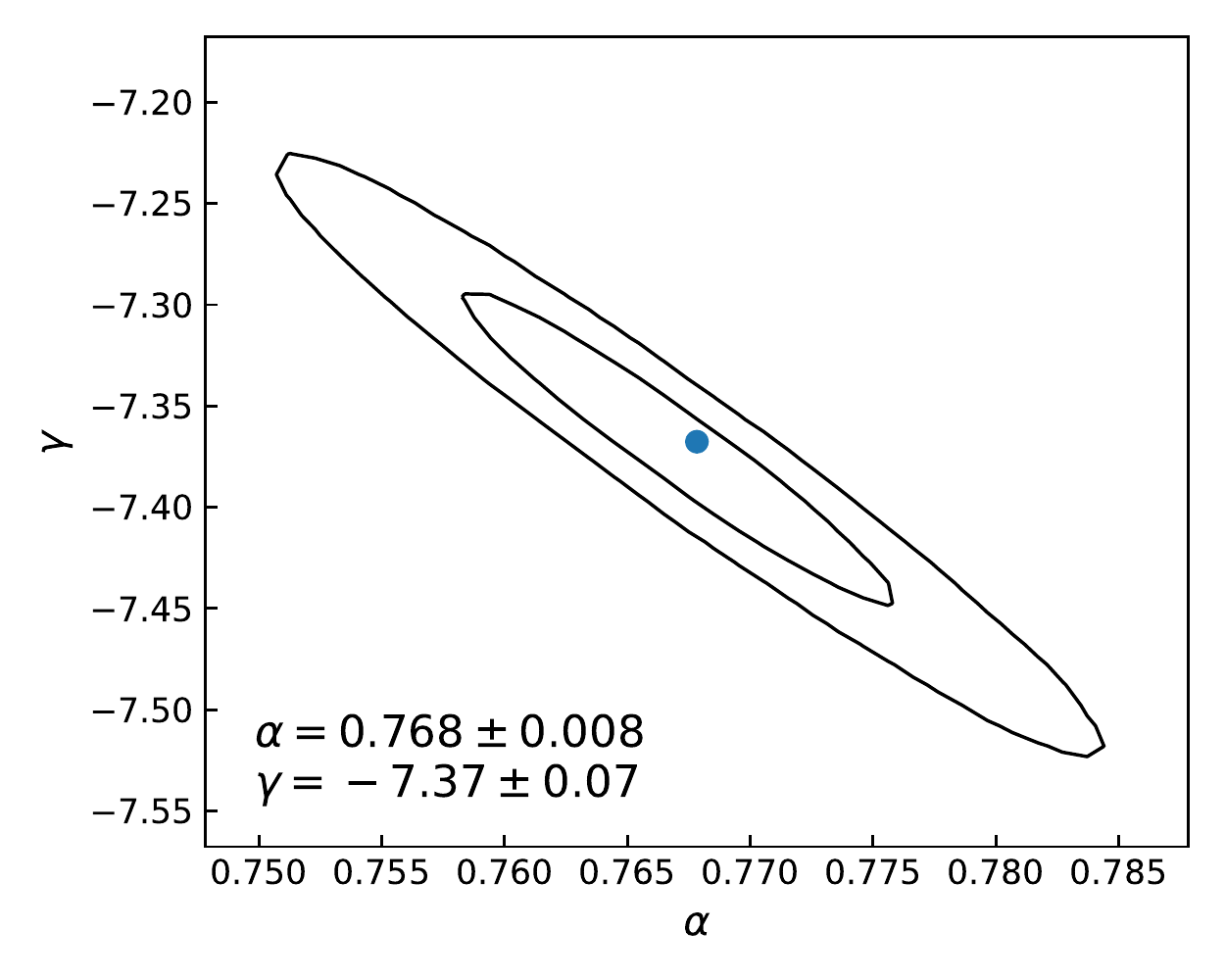}
  \vspace{-.9cm}
  \caption{Confidence contours for linear regression fitting following a redshift-independent SFR--$\Mstar$ relation (Equation~\ref{eqn:fit_noz}).}
  \label{fig:MC_regr_contours_noz}
\end{figure}

%%
%% Figure 9
%%
\begin{figure*}
  \centering
  \includegraphics[width=1\textwidth]{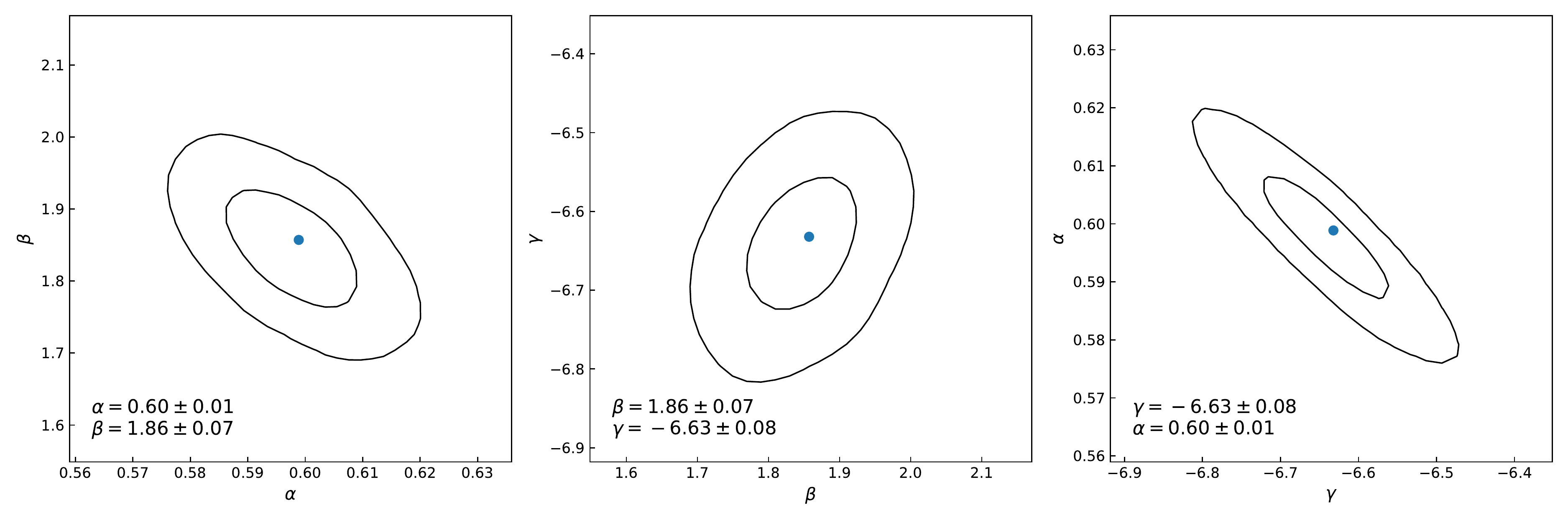}
  \vspace{-.9cm}
  \caption{Similar to Figure~\ref{fig:MC_regr_contours_noz}, but fitting was performed with a redshift-dependent relation (Equation~\ref{eqn:mainseq}).}
  \label{fig:MC_regr_contours}
\end{figure*}

\noindent 
The SFR--$\Mstar$ relation for our sample of \NHaF\ galaxies, with incremental corrections applied (see Section~\ref{sec:NB}--\ref{sec:dust} and Section~\ref{sec:uv_sfrs}), is presented in Figure~\ref{fig:mainseq}.
We find the SFR--$\Mstar$ relation holds down to our low-mass limit cutoff of $10^6 M_\odot$, pushing the explored mass range to $\sim$1.5 dex below previous studies. (The dispersion is $\sigma \approx 0.3$ dex; see Section~\ref{sec:dispersion}.)
For comparison, results
from various studies at $z\lesssim1$ \citep{Noeske07,Salim07,Berg12,Whitaker14,Reyes15} are overlaid.
At the high-mass end, our sample is consistent with \citet{Noeske07} and with the relation of \citet{Salim07}.
The sample from \citet{Berg12} lies consistently below our sample, but this discrepancy may be due to the fact that their galaxies are low-luminosity galaxies in the local Universe, whereas our galaxies are field galaxies at low to intermediate redshifts.
Our relation is below the relation of \citet{Reyes15} and \citet{Whitaker14}.
This discrepancy is likely due to higher sSFR for higher redshift galaxies.

We fit the \Ha\ SFR--$\Mstar$ relation with three different formalisms.
First, we consider a linear relation between mass and SFR:
\begin{equation}
  \log{({\rm SFR})} = \alpha \log{(\Mstar/\Msun)} + \gamma.
  \label{eqn:fit_noz}
\end{equation}
The confidence contours are illustrated in Figure~\ref{fig:MC_regr_contours_noz}, and the
best fit is $\alpha = 0.768 \pm 0.008$ and $\gamma = -7.37 \pm 0.07$, with average residuals of 0.30 dex.

\begin{figure}
  \centering
  \includegraphics[width=1\textwidth]{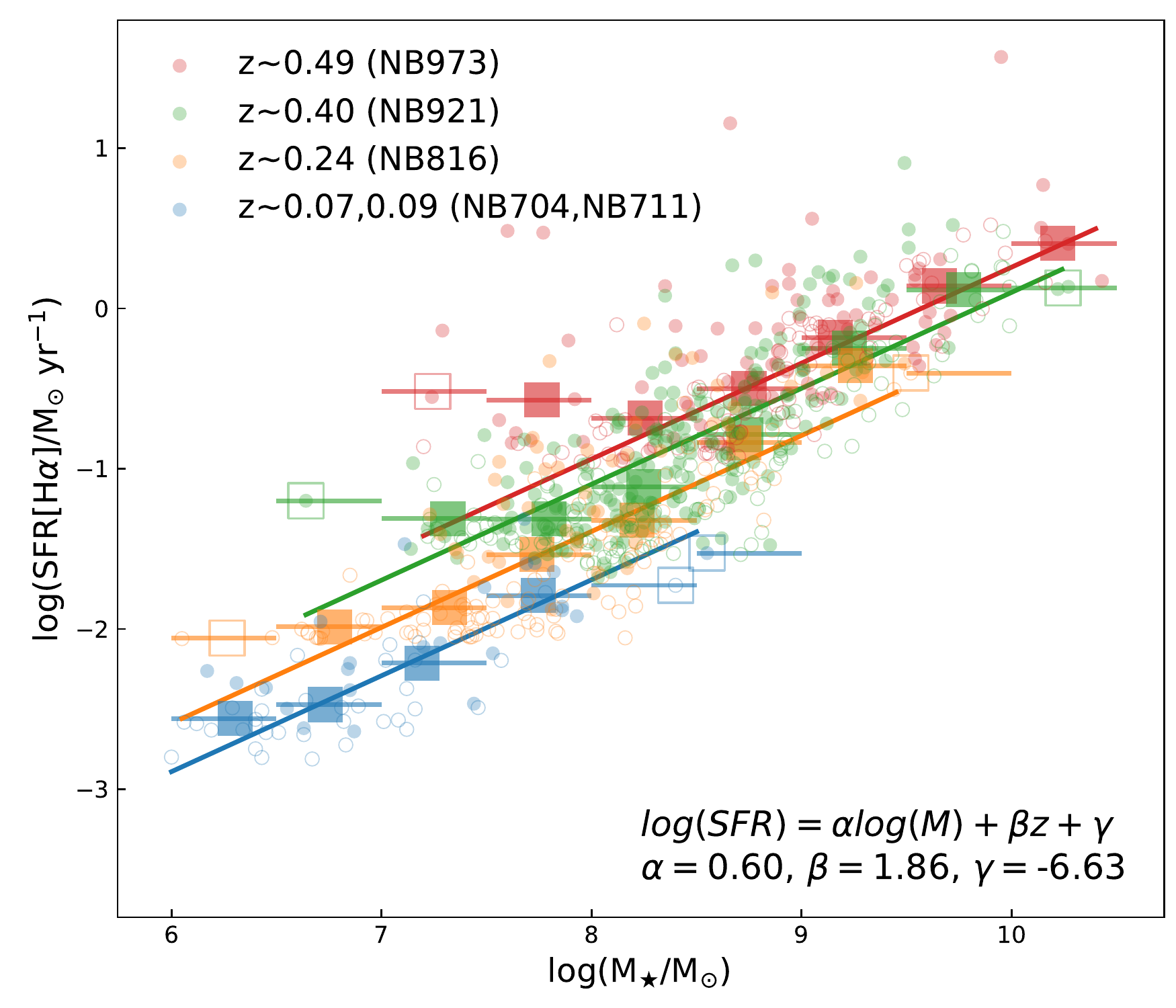}
  \vspace{-0.75cm}
  \caption{The SFR--$\Mstar$ relation with the best fitting redshift-dependent functional form overlaid (solid lines; equation provided in the bottom right). Galaxies are represented by different colors depending on their redshift/NB filter. Empty (filled) circles represent galaxies without (with) spectroscopic confirmation. Squares represent the average SFRs and stellar masses of galaxies in each mass bin, with error bars representing the mass range. Open squares signify there are fewer than five galaxies.}
  \label{fig:zdep}
\end{figure}

Next, we consider a redshift-dependent relation.
We explored various possible redshift-dependent relations, including linear and polynomial dependences on $z$, $(1+z)$, and $\log(1+z)$.
Ultimately, we used:
\begin{equation}
  \label{eqn:mainseq}
  \log{({\rm SFR})} = \alpha \log{(\Mstar/\Msun)} + \beta z + \gamma.
\end{equation}
In this formalism, the redshift dependence in this relation only shifts the normalization of the relationship, and the slope of the stellar mass dependence is fixed for all redshifts. Of the various redshift-dependent relations we explored, this relation had the lowest residuals
(0.26 dex) across stellar mass and redshift, while also having the simplest form. This relation also had lower residuals than the linear, redshift-independent relation considered between mass and SFR.
The best fit with this formalism gives $\alpha = 0.60 \pm 0.01$, $\beta = 1.86 \pm 0.07$, and $\gamma = -6.63 \pm 0.08$. These fit parameters are illustrated in Figure~\ref{fig:MC_regr_contours} with 68\% and 95\% confidence contours. The redshift-dependent SFR--$\Mstar$ relation is provided in Figure~\ref{fig:zdep}. The average mass and SFR values in each stellar mass and redshift bin, provided in Table~\ref{tab:zdep}, agree with the best-fitting redshift dependent model.
Both results indicate a less than unity slope dependence of SFR on mass.

Some studies \citep[e.g.,][]{Whitaker14,Lee15}, have found evidence for a `turnover' at some characteristic mass $M_{\rm char}$. Although their values of $M_{\rm char} \sim 10^{10}~\Msun$ are at the high end of the mass range of our sample, we consider the possibility that there might be a second turnover point at a lower mass range in our sample. As such, we investigate whether a broken power law with redshift dependence gives a better fit:
\begin{equation}
\log({\rm SFR}) = \alpha [\log(\Mstar/M_{\rm char})] + \beta z + \gamma.
\end{equation}
We find that the residuals from considering this broken power law model are similar to those found with the redshift-dependent relation in Equation~\ref{eqn:mainseq}. Given that the redshift-dependent relation has fewer parameters, we decide to use Equation~\ref{eqn:mainseq} as our best-fitting model for the rest of our analysis.

\subsubsection{Comparison against other low-mass SFR--$\Mstar$ relations}
\label{sec:other_studies}
While low-mass galaxy studies have generally been limited, there are two notable studies that do examine the SFR--$\Mstar$ relation in this domain. First, \cite{McGaugh17} examined the SFR--$\Mstar$ relation for low surface brightness galaxies using \Ha\ measurements for SFR. They found that these $z\approx0$ galaxies followed a SFR--$\Mstar$ sequence that is nearly unity ($1.04 \pm 0.06$), much steeper than for more massive star-forming galaxies. However, as discussed in Section~\ref{sec:uv_sfrs}, \Ha\ can systematically underpredict massive star formation. Using the UV--\Ha\ SFR relation of \cite{Lee09}, we find a shallower slope of $0.68 \pm 0.05$. While this slope is higher than our result ($0.60 \pm 0.01$), we note that \cite{McGaugh17} adopted a \cite{Salpeter55} IMF in their \Ha\ SFR estimates instead of a \cite{Chabrier03}. Given the difference in IMF, the $2\sigma$ offset is marginal.

Second, \cite{Boogaard18} examined low-mass galaxies, $ 7 \leq \log(M_\star / M_\odot) \leq 10.5 $, in the {\it Hubble} Deep Field using data from the MUSE \citep[Multi Unit Spectroscopic Explorer;][]{Bacon10} instrument.
They looked at 179 star-forming galaxies at redshift $ 0.11 < z < 0.91$ with \Hb\ and \Ha\ emission lines (with high spectroscopic measurements), from which SFRs were determined. 
The SFR--$\Mstar$ relation they found was $\alpha' = 0.83^{+0.07}_{-0.06}$ and $\beta' = 1.74^{+0.66}_{-0.68}$ for the functional form ${\log(\rm{SFR}) \propto \alpha' \log(M_\star/M_\odot) + \beta' \log(1+z)}$. Notably, their value of $ \alpha' = 0.83 $ significantly differs from our value of $ \alpha = 0.58$.
However, there are a few differences in the way our analyses were conducted:
(1) their value of $ \alpha' $ for their redshift-dependent relation spanned a redshift range beyond our own; (2) their high-$z$ sample was dominated by \Hb-selected galaxies; and (3) they also did not apply the UV correction as we did.

In order to more directly compare our results, we examine the 72 galaxies for $ 0.1 < z \leq 0.5 $ from
\cite{Boogaard18}\footnote{Data provided directly by Boogaard et al.}, and derive SFRs in a consistent manner.
We find that our samples overlap significantly across the full stellar mass range.
We note however, that the fitting methods used differ, and thus differences in the fitting results are expected.

%%
%% Figure 10
%%

\subsection{Intrinsic Dispersion of the Star Formation Sequence Relation}
\label{sec:dispersion}
\begin{figure}
  \centering
  \includegraphics[width=1\textwidth]{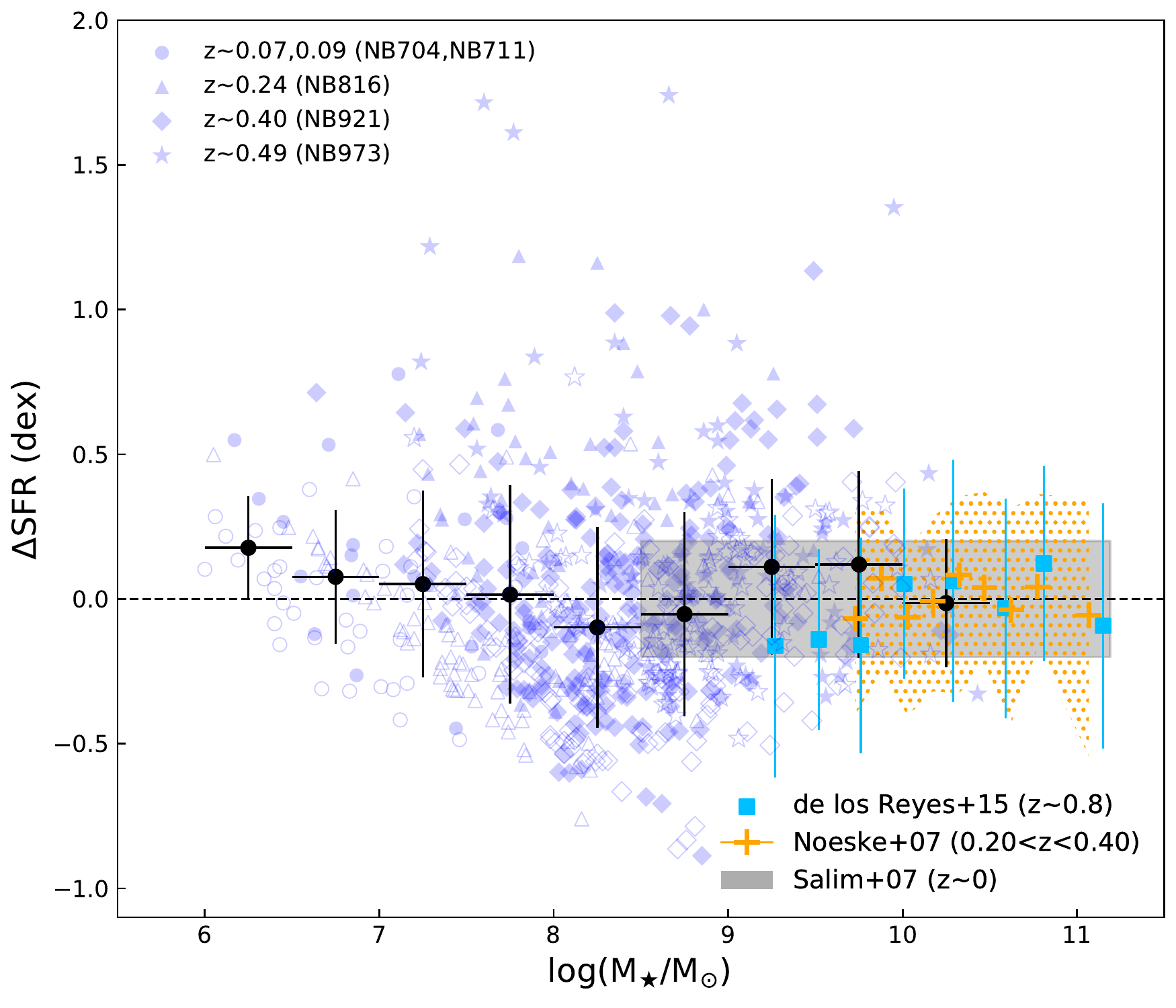}
  \vspace{-0.7cm}
  \caption{Deviations from the SFR--$\Mstar$ relation (relative to the redshift-dependent SFR--$\Mstar$ relation, Equation~\ref{eqn:mainseq}) for \Ha\ emitters between $z=0.07$ and $z=0.5$.
  	Empty (filled) symbols represent galaxies without (with) spectroscopic confirmation.
  	Vertical error bars for filled black circles illustrate the average dispersion in each bin, with horizontal error bars indicating the mass range. Results from other empirical studies are overlaid: \citet{Reyes15} with cyan squares, \citet{Noeske07} with orange pluses, and \citet{Salim07} with a gray shaded region.}
  \label{fig:mainseq_dispersion}
\end{figure}

The residuals of the SFR--$\Mstar$ relation, illustrated in Figure~\ref{fig:mainseq_dispersion}, are obtained by differencing the dust-corrected SFR values against those predicted by the redshift-dependent SFR--$\Mstar$ relation (Equation~\ref{eqn:mainseq}). The sample is divided into stellar mass bins, and the observed scatter is illustrated in this figure. Each mass bin also has an associated uncertainty due to systematics, which is derived from adding in quadrature the errors on the NB fluxes and on the \EBVm\ values. Subtracting in quadrature the systematic uncertainties from the observed scatter yields the estimated intrinsic dispersion as a function of mass for our SFR--$\Mstar$ relation. These values are provided in Table~\ref{tab:dispersion}.

%%
%% Figure 11
%%

Over the stellar mass range of 10$^{7.0}$--10$^{8.5}~\Msun$, we observe that the intrinsic dispersion varies between $\approx$0.25--0.35 dex, with an average of $\approx$0.3 dex.\footnote{Incompleteness is a large factor for dispersion calculations below $10^7~\Msun$.} The average intrinsic dispersion across the entire mass range is also $\approx$0.3 dex, suggesting there is no evolution of dispersion towards lower stellar masses.
This result is aligned with similar intrinsic dispersion measures found in other studies probing higher mass ranges \citep[e.g.,][]{Noeske07, Whitaker12b, Speagle14, Schreiber15}.

Over similar stellar masses, our intrinsic dispersion is consistent with \citet{McGaugh17}, where they find $\sigma = 0.34$ dex.
However, our intrinsic dispersion of $\approx$0.3 dex differs from \citet{Boogaard18}, where they found an intrinsic dispersion of $0.44$ dex. One reason for the difference is that our study, as well as other studies mentioned above, are affected by the selection effects that miss low-sSFR galaxies. With our mock galaxy simulations (see Section \ref{sec:comp}), we find that the dispersion can be underestimated by a factor of $\approx$2 such that the SFR--$\Mstar$ dispersion is $\approx$0.5 dex.  Our simulations show a consistent dispersion across the stellar mass range 10$^6$--10$^{10}$ $\Msun$, which suggests that there does not appear to be a significant evolution with stellar mass.
The higher dispersion is consistent with local studies, such as \citet{Davies19}, where they find a dispersion of $\approx$0.4 dex above $3\times10^8$ $\Msun$ with a slow rise toward $\approx$0.6 dex at 10$^7$ $\Msun$.

\begin{table}[ht]
  \centering
  \caption{Summary of dispersion measurements from the SFR--$\Mstar$ relation}
  \label{tab:dispersion}
  \begin{threeparttable}
    \centering
    \begin{tabular}{lccccc}
      \hline\hline
      $\log\left(\frac{\Mstar}{\Msun}\right)$ &
      $\left<\log{\left(\frac{{\rm SFR}}{\Msun/{\rm yr}}\right)}\right>$ &
      $N$ & $\sigma_{\rm obs}$ & $\left<\sigma_{\rm sys}\right>$ & $\left<\sigma_{\rm int}\right>$ \\
      (1) & (2) & (3) & (4) & (5)  & (6) \\
      \hline

6.0--6.5   & --2.50 (--3.02) &  18 & 0.178 & 0.247 & \ldots \\
6.5--7.0   & --2.23 (--2.65) &  29 & 0.230 & 0.197 & 0.119 \\
7.0--7.5   & --1.74 (--2.02) &  68 & 0.322 & 0.202 & 0.251 \\
7.5--8.0   & --1.35 (--1.54) & 127 & 0.378 & 0.190 & 0.326 \\
8.0--8.5   & --1.10 (--1.27) & 170 & 0.347 & 0.143 & 0.316 \\
8.5--9.0   & --0.68 (--0.85) & 164 & 0.354 & 0.112 & 0.336 \\
9.0--9.5   & --0.23 (--0.40) &  88 & 0.302 & 0.053 & 0.297 \\
9.5--10.0  &  +0.12 (--0.05) &  42 & 0.322 & 0.035 & 0.320 \\
10.0--10.5 &  +0.34 (+0.17)  &   8 & 0.222 & 0.024 & 0.220 \\
	
      \hline
    \end{tabular}
    \begin{tablenotes}
    \item \leavevmode\kern-\scriptspace\kern-\labelsep (1): Stellar mass range. (2): Average log SFR (average log SFR without FUV corrections). (3): Number of galaxies in each bin. (4)--(6): Observed scatter, average systematic uncertainties, and average intrinsic dispersion on the main sequence relation, respectively. These values are illustrated in Table~\ref{fig:mainseq_dispersion}.
  \end{tablenotes}
  \end{threeparttable}
\end{table}

\subsection{Survey Selection and Completeness}
\label{sec:comp}
The NB technique selects galaxies with (1) a minimum excess in the NB color (corresponding to a minimum observed emission-line EW) at the bright end; and (2) a 3$\sigma$ NB excess flux limit. These observable limits are provided in Table~\ref{tab:survey_selection} for all five NB filters.

To understand the completeness of our study, we conduct Monte Carlo analyses where mock emission-line galaxies are created and selected through our NB criteria. This approach allows us to examine the selection effects across a number of observables including stellar mass, emission-line luminosity/flux, and emission-line EW. Our methodology is similar to \cite{Ly11b} where we first adopt log-normal EW distributions, which are described by two quantities: the mean EW, $\left<\log{\rm EW}\right>$, and the dispersion, $\sigma[\log({\rm EW})]$.  We chose a log-normal distribution because a number of studies have found that the \Ha\ emission-line EWs follow such a distribution \citep[e.g.,][]{Lee07}.

\begin{figure*}
  \centering
  \includegraphics[width=0.45\textwidth]{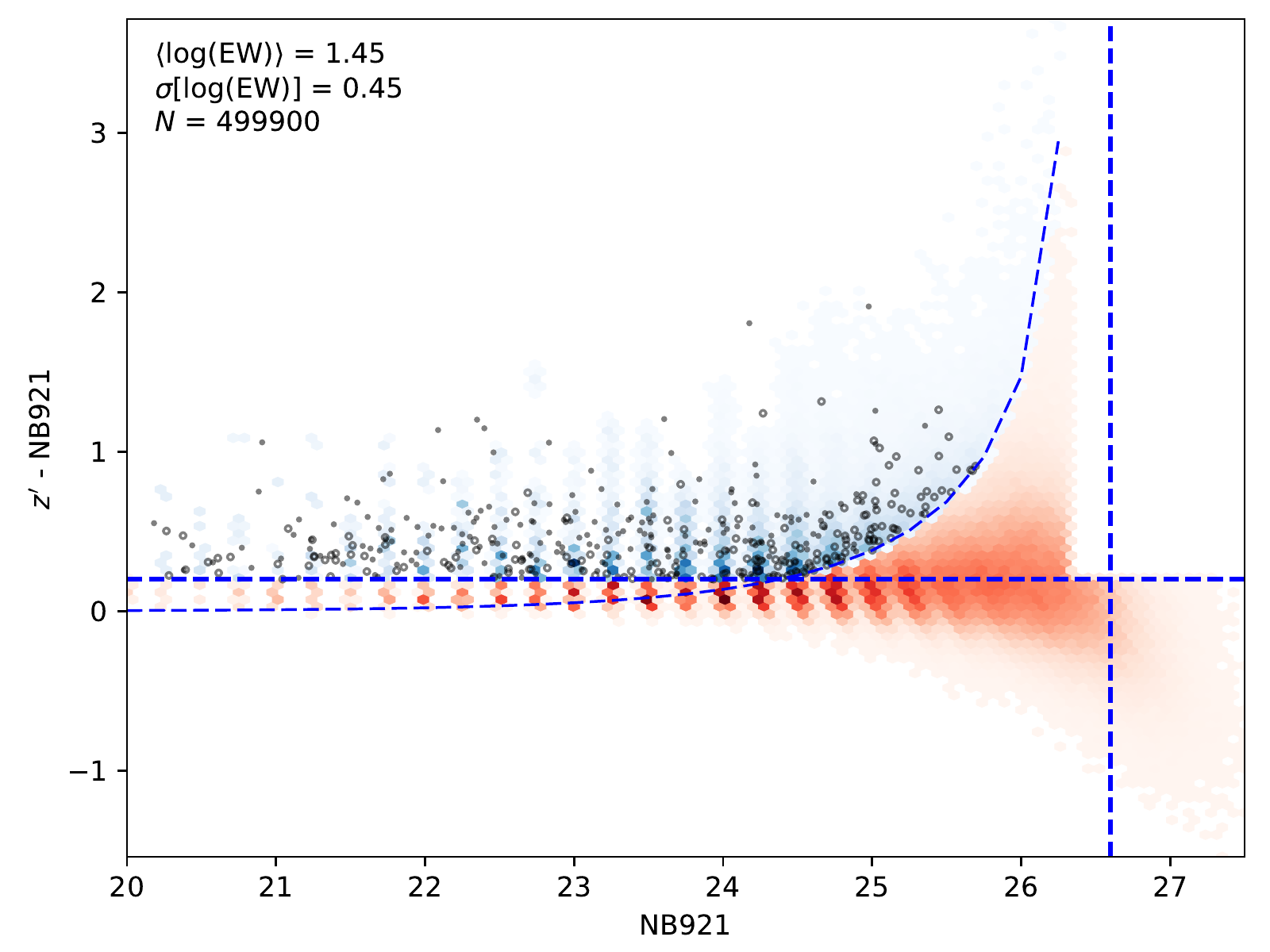}
  \includegraphics[width=0.45\textwidth]{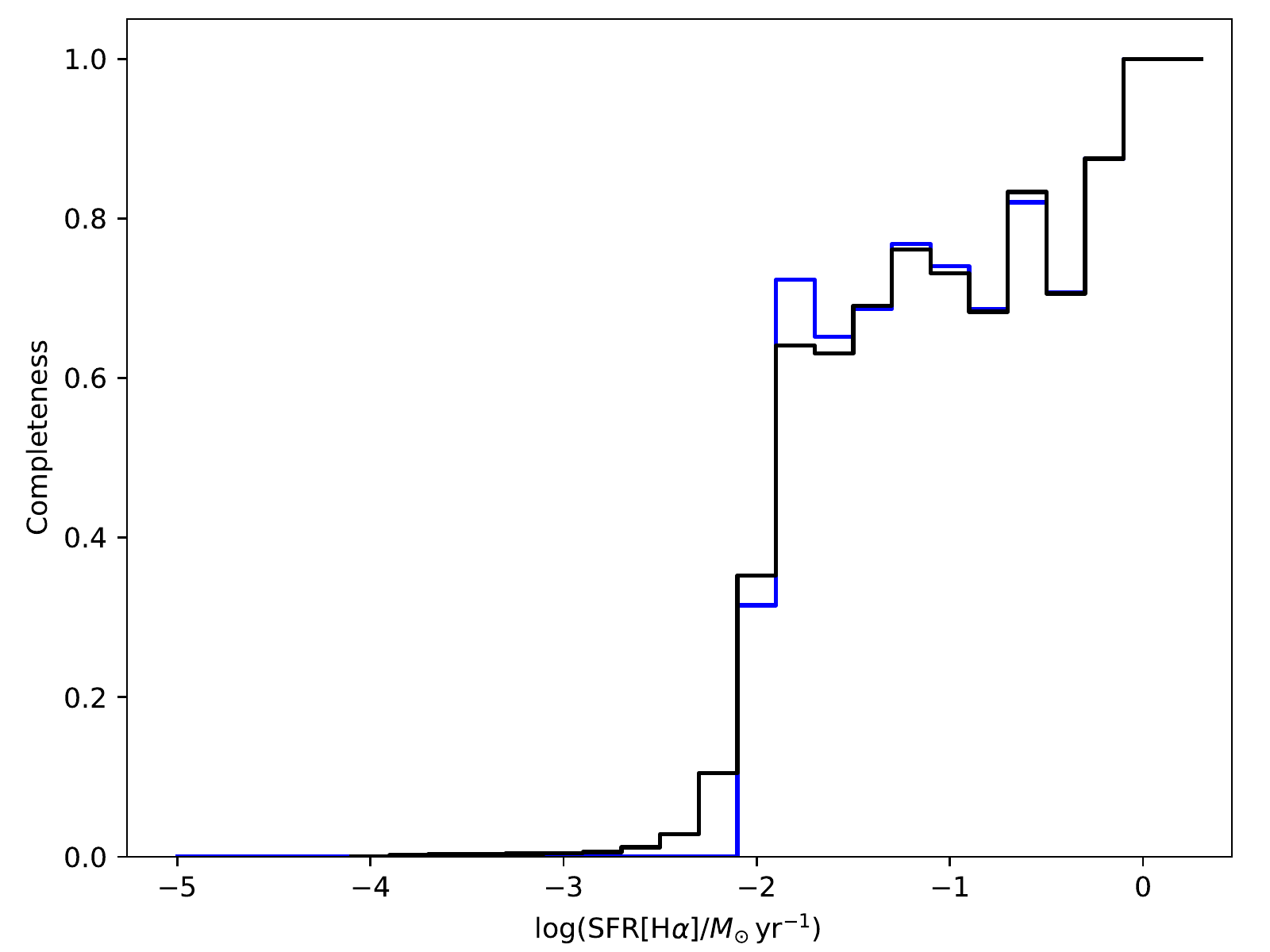}
  \vspace{-.3cm}
  \caption{NB selection diagram (left panel) and completeness limit for SFR (right panel) for NB921 excess emitters using the best fit model, $\left<\log{\rm EW}\right>=1.45$ and  $\sigma[\log({\rm EW})]=0.45$.
    Blue dashed lines indicate our NB excess selection with the vertical dashed line indicating the NB921 3$\sigma$
    limit. Blue (red) points indicate galaxies that meet (miss) our selection, with color intensity indicating
    more galaxies. The right panel illustrates that our selection reaches a plateau at high SFRs due to
    missing low-sSFR galaxies. It then follows a monotonic decrease with lower SFRs. The black (blue) line is based
    on modeled (measured) SFRs.}
  \label{fig:comp}
\end{figure*}

For our Monte Carlo analysis, we consider a grid of $\left<\log{\rm EW}\right>$ and $\sigma[\log({\rm EW})]$ values, a total of 16 possibilities for each NB filter selection.  In each simulation, we assign NB magnitudes for 5,000 modeled galaxies. The magnitudes of these sources are governed by the number counts obtained from the SDF NB SExtractor catalogs \citep{Ly07,Ly12a,Ly12b}. 
By adopting a given log-normal EW distribution, a NB excess EW/color can be assigned to each mock galaxy. This ultimately populates the NB excess selection plot of NB excess color vs. NB magnitude (see Figure~\ref{fig:comp}). For each modeled galaxy, we construct 100 randomizations with photometric uncertainties governed by limiting magnitudes for NB and broadband data. We then apply our NB color selections and identify which mock galaxies are selected. With mock photometry, we determine derived properties such as the emission-line flux and EW from the NB excess. We also apply corrections for the NB filter shape (see Section~\ref{sec:NB}) and \NII\ contamination (see Section~\ref{sec:NII}) to derive \Ha\ emission-line flux/luminosity and EW. To determine stellar masses for each mock galaxy, we construct an empirical relationship between broadband photometry and stellar mass estimates for our NB emitter samples.  For simplicity, we use the broad-band filters that are used in our NB excess selection. We illustrate the derived properties of our mock galaxies in Figure~\ref{fig:comp}.

With 16 possible log-normal EW distribution each with about 500,000 mock galaxies, and five NB filters, a total of 40 million galaxies were constructed to ensure that our simulations cover a wide range of possible priors. To determine which log-normal distribution represented the sample of emission-line galaxies selected in each filter, we compare the distributions of EW and line fluxes against what is measured from the observable samples. Here we normalise the mock distributions against the sample sizes for each filter, and computed reduced $\chi^2$ values. The best log-normal EW model for each NB filter was determined by using a weighted $\chi^2$ for both EW and line flux. In Table~\ref{tab:survey_selection} we report the survey completeness at the 50\% level for sSFR and SFR. Here the sSFR (SFR) completeness limit is at higher (lower) luminosities, where the minimum EW (NB~3$\sigma$ excess) determines our selection.

With these Monte Carlo simulations, we can determine how our adopted NB color selection impacts the SFR--$\Mstar$ relation and %effects
our results. These color selections preferentially will miss low-mass low-SFR galaxies, which has two immediate effects on the SFR--$\Mstar$ relation. First the slope can be flattened. Second, the amount of dispersion will be underestimated. We examine these effects in the following manner.
We first compare the observed slope (after applying the NB selection) against the intrinsic slope, limiting our analysis to the best-fitting log-normal EW models that we describe above for each filter. We find that the slope is not significantly impacted. The slope will not be notably flattened because low-EW galaxies are not selected across all stellar masses.

For the dispersion, we (1) randomly select, from our mock samples, the number of $\mathcal{MACT}$ galaxies in each $\Mstar$--$z$ bin, (2) compute the offset relative to the best-fitting SFR--$\Mstar$ relation for each redshift, and then (3) combine the results of each NB filter to determine the standard deviation. We repeat this computation 1,000 times to estimate the range in dispersion, and then compare our results to the intrinsic dispersion, which is measured in the SFR--$\Mstar$ relation without any NB color selection applied. We find that the NB-selected SFR--$\Mstar$ relation under-predicts the SFR dispersion by a factor of $\approx$2 over the stellar mass range of $10^6$--$10^{10}$ $\Msun$, suggesting that the dispersion in the SFR--$\Mstar$ relation is $\approx$0.5 dex.

%%
%% Figure 12
%%

\begin{table*}
  \centering
  \caption{Summary of NB Selections}
  \label{tab:survey_selection}
  \begin{threeparttable}
    \begin{tabular}{lcccccccc}
    \hline\hline
    Filter & (NB--BB)$_{\rm min}$ & EW$_{\rm obs, min}$ & EW$_{\rm rest, min}$ & $\log{F_{\rm NB}}$ &
    $\left<\log({\rm EW})\right>$ & $\sigma[\log({\rm EW})]$ & $C$(sSFR) & $C$(SFR[\Ha])\\
    (1)    & (2)                 & (3)               &       (4)          & (5) & (6) & (7) &  (8) & (9) \\
    \hline
    NB704 & 0.15 & 16.5 & 15.4 & -17.26 & 1.25 & 0.55 & -10.22 & -3.66 \\
    NB711 & 0.15 & 11.5 & 10.6 & -17.20 & 1.05 & 0.75 & -10.20 & -3.45 \\
    NB816 & 0.15 & 19.7 & 15.9 & -17.36 & 1.25 & 0.45 & -10.19 & -2.57 \\
    NB921 & 0.20 & 32.0 & 22.8 & -17.23 & 1.45 & 0.45 & -10.06 & -1.78 \\
    NB973 & 0.25 & 70.3 & 47.3 & -16.77 & 1.25 & 0.45 &  -9.69 & -0.09 \\
    \hline
    \end{tabular}
    \begin{tablenotes}
    \item \leavevmode\kern-\scriptspace\kern-\labelsep (1) NB filter. (2) Minimum NB color excess in magnitudes. (3) Minimum observed \Ha+\NII\ EW in Angstroms. (4) Minimum rest-frame \Ha+\NII\ EW in Angstroms. (5) Logarithm of 3-$\sigma$ NB excess flux limit in erg s$^{-1}$ cm$^{-2}$. (6)--(7) Best log-normal EW distribution model described with mean and dispersion. (8)--(9) Completeness limit at 50\% for $\log\left({\rm sSFR/yr}^{-1}\right)$ and $\log\left(\frac{{\rm SFR[H}\alpha]}{\Msun~{\rm yr}^{-1}}\right)$.
    \end{tablenotes}
  \end{threeparttable}
\end{table*}

\section{Conclusions}
\label{sec:concl}
\noindent
With our sample of \NHaF\ \Ha\ emitting NB-selected galaxies from the \textit{Metal Abundances across Cosmic Time} ($\mathcal{MACT}$) Survey, we have been able to observationally extend the ``star formation main sequence" and the specific SFR evolution down to 10$^6 \Msun$, a domain that had not yet been explored with observed individual galaxies outside of the local Universe ($ z \gtrsim 0.07 $). Our key results are as follows:
\begin{enumerate}[label=\arabic*.,leftmargin=1\parindent] 
\item Our \Ha-derived SFRs are systematically underpredicted relative to FUV-derived SFRs, in agreement with observations of local galaxies \citep[e.g.,][]{Lee09}.
\item At a given stellar mass, $\log(\Mstar/\Msun) = 8.0 \pm 0.5$, there is significant evolution for the specific SFR: ${\log \left(\rm sSFR\right) \propto A \log(1+z)}$ with $A = 5.26\pm 0.75$. On average, low-mass galaxies have higher sSFR than more massive ($3\times10^9 \Msun$) galaxies, with the sSFR nearly reaching a plateau at low stellar masses.
\item The SFR--$\Mstar$ relation holds for galaxies spanning $\sim$10$^6$--$10^{10}~\Msun$ at $z \approx$ 0.07--0.5, and follows a redshift-dependent relation of $\log{({\rm SFR})} \propto \alpha \log(\Mstar/\Msun) + \beta z$ with $\alpha = 0.60 \pm 0.01$ and $\beta = 1.86 \pm 0.07$. Our results do not significantly differ from recent studies that have also explored the SFR--$\Mstar$ relation for low-mass galaxies, albeit across slightly different redshift ranges \citep[e.g.,][]{McGaugh17, Boogaard18}.
\item The observed intrinsic dispersion of our galaxies appears to be fairly well regulated ($\approx$0.3 dex), in agreement with other observational work at comparable redshifts. With corrections for survey selection effects that miss low specific SFR galaxies, we find that our observed intrinsic dispersion may be underestimated by a factor of $\approx$2 such that the true dispersion is $\approx$0.5 dex. We do not observe any evidence for increasing or decreasing scatter over our mass range, 10$^6$--10$^{10}~\Msun$, which is consistent with results obtained at similar or higher masses.
\end{enumerate}

\section*{Acknowledgements}

We thank the anonymous referee for helpful feedback that improved the quality of this paper.
This work was supported by a NASA Keck PI Data Award, administered by the NASA Exoplanet Science Institute. Data presented herein were obtained at the W. M. Keck Observatory from telescope time allocated to the National Aeronautics and Space Administration through the agency's scientific partnership with the California Institute of Technology and the University of California. The Observatory was made possible by the generous financial support of the W. M. Keck Foundation. The authors wish to recognize and acknowledge the very significant cultural role and reverence that the summit of Mauna Kea has always had within the indigenous Hawaiian community. We are most fortunate to have the opportunity to conduct observations from this mountain.
Hectospec observations reported here were obtained at the MMT Observatory, a joint facility of the Smithsonian Institution and the University of Arizona. A subset of the MMT telescope time was granted by NOAO, through the NSF-funded Telescope System Instrumentation Program (TSIP). We thank Perry Berlind, Michael Calkins, and Ben Weiner for acquisition of the Hectospec data. We thank Nobunari Kashikawa for help in acquisition of Keck data.
We thank Leindert Boogaard for sharing data to compare analyses and helpful private communications.
We also thank Bruce Macintosh for his support.
We gratefully acknowledge NASA's support for construction, operation, and science analysis for the {\it GALEX} mission.
This research made use of the following \textsc{python} packages: \texttt{numpy} \citep{Oliphant06, Harris20}, \texttt{scipy} \citep{Jones01, Virtanen20}, \texttt{matplotlib} \citep{Hunter07}, and \texttt{astropy} \citep{Astropy13, Astropy18}.

\section*{Data availability}
The data in this article are available upon request to the corresponding author.

%%%%%%%%%%%%%%%%%%%%%%%%%%%%%%%%%%%%%%%%%%%%%%%%%%



%\bibliographystyle{mnras}


%%%%%%%%%%%%%%%%%%%%%%%%%%%%%%%%%%%%%%%%%%%%%%%%%%

%%%%%%%%%%%%%%%%% APPENDICES %%%%%%%%%%%%%%%%%%%%%

\appendix

\section{Color selection plots}

In this appendix, we illustrate the broad-band color selections that were adopted to identify \Ha\ emitters.
We refer to Section \ref{sec:ha_nb_emits} for more details on the color selections and the \Ha\ samples.

\begin{figure*}
  \centering
  \includegraphics[width=1.0\textwidth]{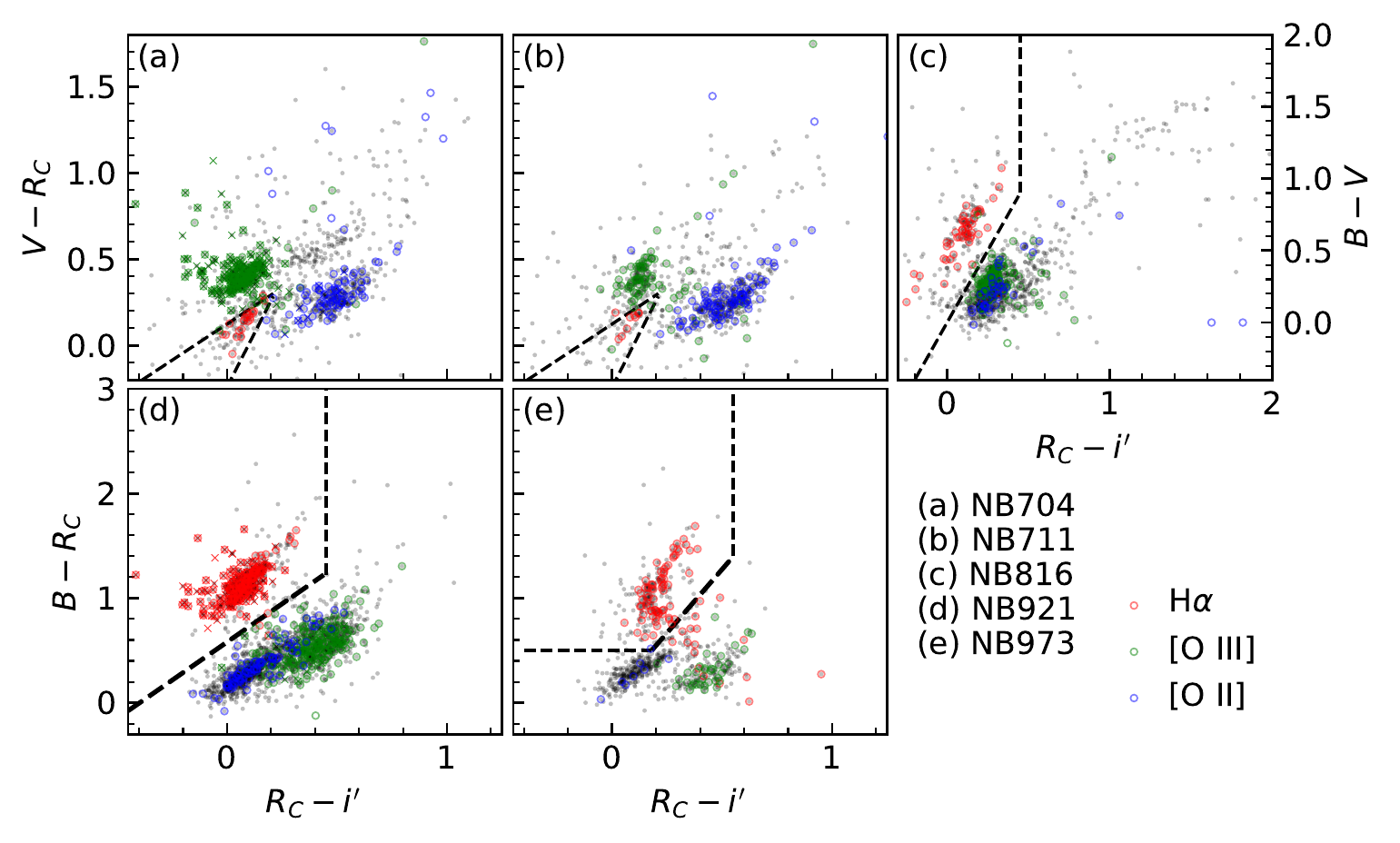}
  \vspace{-0.5cm}
  \caption{
    Two color selection of \Ha\ excess emitters in (a) NB704, (b) NB711, (c) NB816, (d) NB921, and (e) NB973
        filters. The $x$ axes illustrate the $\Rc-i^{\prime}$ colors while the $y$ axes illustrate $V-\Rc$ (NB704,
        and NB711), $B-V$ (NB816), and $B-\Rc$ (NB921 and NB973). Galaxies with spectroscopic confirmation are
        indicated with open circles with colors indicating NB emission line: \Ha\ (red), \OIII+\Hb\ (green), and
        \OII\ (blue). In addition, dual NB704+NB921 emitters are overlaid as crosses, which occurs for two reasons:
        (1) \OIII\ and \Ha\ (red crosses), and (2) \OII\ and \Hb\ (blue crosses), in NB704 and NB921 filters,
        respectively.}
  \label{fig:color_selection}
\end{figure*}

\section{Summary of composite spectra measurements}

In this appendix, we provide figures that illustrate our composite spectra, as well as tables that summarise our results.
We refer to Section \ref{sec:stack} for more details on the composite spectral stacking and emission-line fitting.

%%
%% Figure B1
%%
\begin{figure*}
	\centering
	\includegraphics[width=0.92\textwidth]{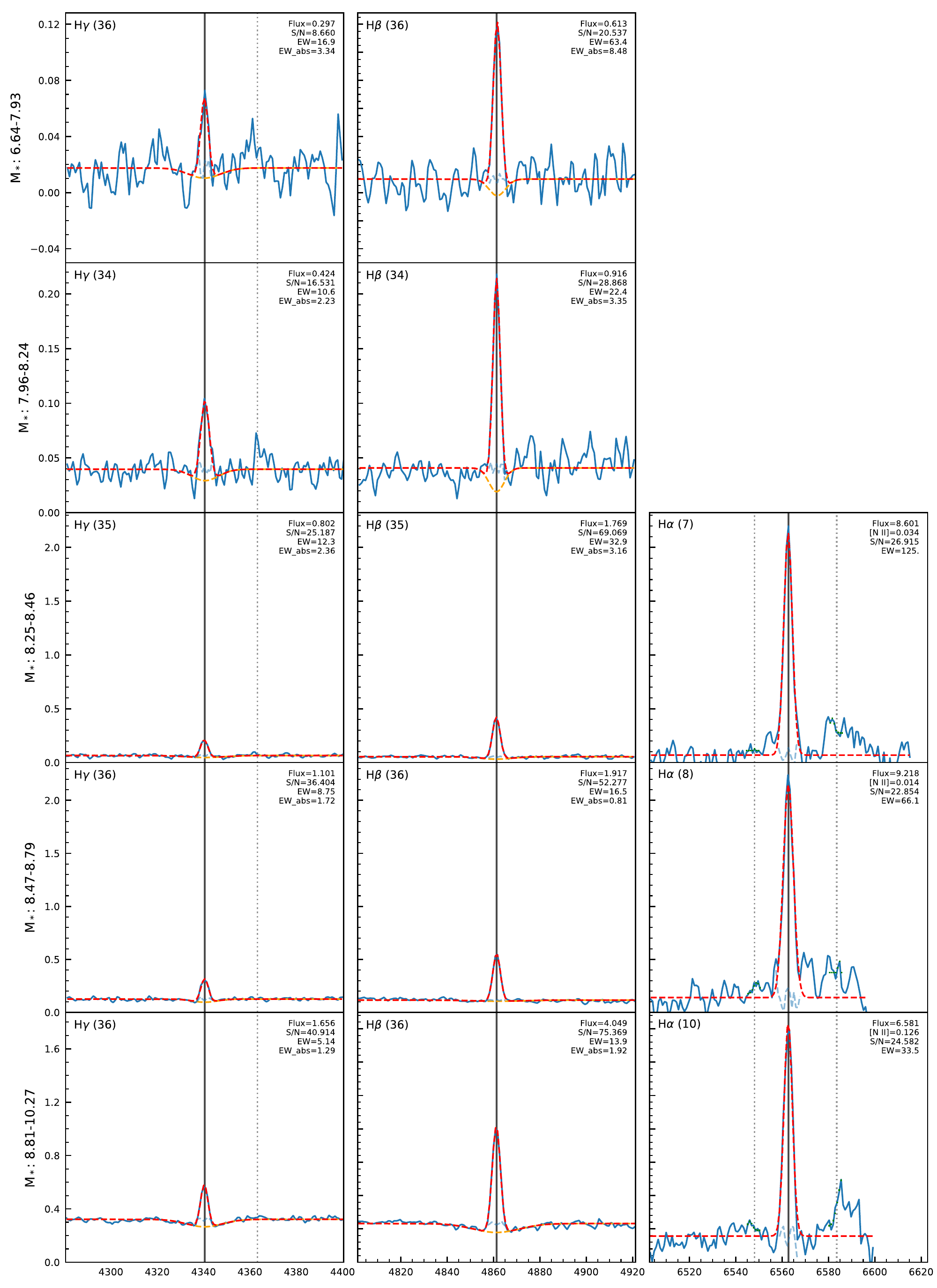}
	\vspace{-0.3cm}
	\caption{Composite spectra for the MMT NB921 sample (solid blue) with the best-fitting Gaussian profile (dashed red) and the RMS (dashed light blue). For the left and middle panels, the best-fitting Balmer absorption Gaussian profile is overlaid as well (dashed orange). This sample is split into five $\Mstar$ bins. The $x$-axis is rest-frame wavelength in units of \AA\ and the $y$-axis is flux density in units of 10$^{-17}$ erg s$^{-1}$ cm$^{-2}$ \AA$^{-1}$. \NIIl\ lines are denoted by the vertical dotted lines in the right-most panels. ``\NII'' in the upper right legend denotes the emission-line flux ratio of \NII$\lambda$6583 to \Ha. [\textsc{O~iii}]$\lambda$4363 lines are denoted by the vertical dotted lines in the left-most panels.}
	\label{fig:stacked_gals_mmt}
\end{figure*}

%%
%% Figure B2
%%
\begin{figure}
	\centering
	\includegraphics[width=1\textwidth]{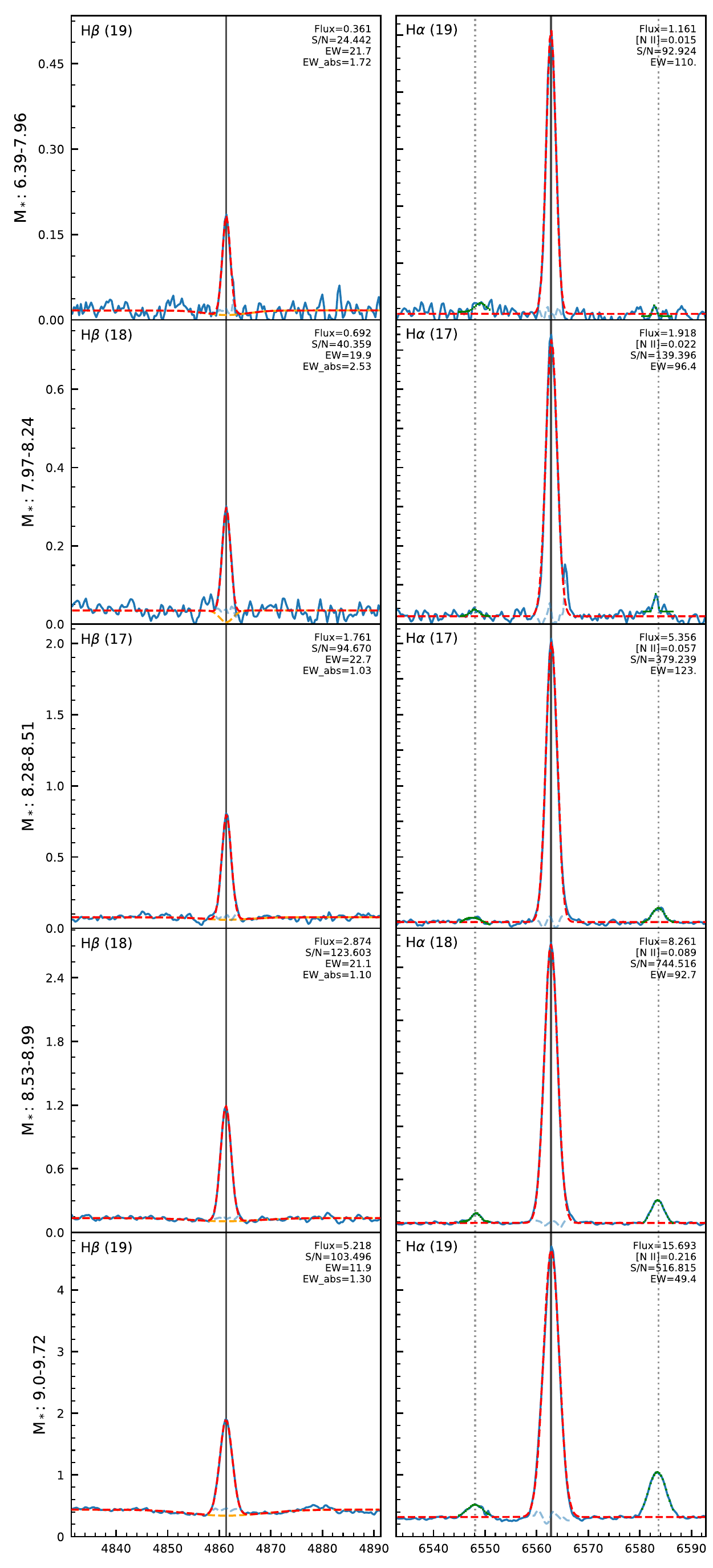}
	\vspace{-.7cm}
	\caption{
		Similar to Figure~\ref{fig:stacked_gals_mmt}, but for the Keck NB921 sample.
	}
  \label{fig:stacked_gals_keck}
\end{figure}

\begin{table*}[ht]
  \centering
  \caption{Summary of measurements from MMT/Hectospec composite spectra}
  \label{tab:mmt_stacked_gals}
  \begin{threeparttable}
    \centering
    \begin{tabular}{lccccccccccc}
      \hline\hline
      $\log(\Mstar/\Msun)$ & $N_{\rm H\gamma}$ & $N_{\rm H\beta}$ & $N_{\rm H\alpha}$ & $F_{\rm H\beta}$ & $F_{\rm H\gamma}$/$F_{\rm H\beta}$ & $F_{\rm H\alpha}$/$F_{\rm H\beta}$ & \EBVm$_{\rm H\gamma}$ & \EBVm$_{\rm H\alpha}$ & $F_{[\textsc{N~ii}]}$/$F_{\rm H\alpha}$ \\
      (1) & (2) & (3) & (4) & (5) & (6) & (7) & (8) & (9) & (10)\\
      \hline
      \multicolumn{10}{c}{NB704+NB711 ($z=0.068$--0.091)}\\
6.17--7.20 & 12 & 12 & 12 & 34.84$_{-0.54}^{+0.57}$ & 0.50$_{-0.01}^{+0.01}$ & 2.15$_{-0.01}^{+0.01}$ & 0.000$_{-0.035}^{+0.037}$ & 0.000$_{-0.006}^{+0.007}$ & 0.14 \\
7.28--8.55 & 12 & 12 & 12 & 55.25$_{-1.18}^{+1.25}$ & 0.28$_{-0.02}^{+0.01}$ & 2.40$_{-0.04}^{+0.04}$ & 0.988$_{-0.098}^{+0.107}$ & 0.000$_{-0.016}^{+0.016}$ & 0.04 \\
	\noalign{\vskip 1mm}

      \hline
      \multicolumn{10}{c}{NB816 ($z=0.229$--0.252)}\\
7.23--7.60 & 10 & 10 & 10 & 16.15$_{-0.67}^{+0.70}$ & 0.40$_{-0.03}^{+0.03}$ & 2.25$_{-0.04}^{+0.04}$ & 0.285$_{-0.133}^{+0.157}$ & 0.000$_{-0.017}^{+0.019}$ & 0.08 \\
7.70--7.93 & 10 & 10 & 10 & 35.61$_{-0.66}^{+0.70}$ & 0.44$_{-0.01}^{+0.01}$ & 2.04$_{-0.01}^{+0.01}$ & 0.133$_{-0.037}^{+0.039}$ & 0.000$_{-0.006}^{+0.006}$ & 0.08 \\
7.97--8.17 & 10 & 10 & 10 & 23.00$_{-0.55}^{+0.53}$ & 0.31$_{-0.02}^{+0.02}$ & 2.10$_{-0.01}^{+0.01}$ & 0.815$_{-0.107}^{+0.122}$ & 0.000$_{-0.004}^{+0.005}$ & 0.01 \\
8.21--8.60 & 9 & 9 & 9 & 77.13$_{-0.66}^{+0.60}$ & 0.37$_{-0.00}^{+0.00}$ & 2.16$_{-0.01}^{+0.01}$ & 0.470$_{-0.024}^{+0.024}$ & 0.000$_{-0.004}^{+0.004}$ & 0.10 \\
8.65--9.49 & 10 & 10 & 10 & 60.97$_{-1.03}^{+0.87}$ & 0.34$_{-0.01}^{+0.01}$ & 2.54$_{-0.02}^{+0.02}$ & 0.618$_{-0.056}^{+0.055}$ & 0.000$_{-0.007}^{+0.007}$ & 0.54 \\
	\noalign{\vskip 1mm}
      
      \hline
      \multicolumn{10}{c}{NB921 ($z=0.389$--0.413)}\\
6.64--7.93 & 36 & 36 & 1 & 6.13$_{-0.28}^{+0.30}$ & 0.48$_{-0.03}^{+0.03}$ & \ldots & 0.000$_{-0.111}^{+0.131}$ & \ldots & \ldots \\
7.96--8.24 & 34 & 34 & \ldots & 9.16$_{-0.31}^{+0.33}$ & 0.46$_{-0.01}^{+0.01}$ & \ldots & 0.022$_{-0.046}^{+0.051}$ & \ldots & \ldots \\
8.25--8.46 & 35 & 35 & 7 & 17.69$_{-0.26}^{+0.25}$ & 0.45$_{-0.01}^{+0.01}$ & 1.69$_{-0.05}^{+0.05}$ & 0.062$_{-0.049}^{+0.052}$ & 0.000$_{-0.031}^{+0.030}$ & 0.05 \\
8.47--8.79 & 36 & 36 & 8 & 19.17$_{-0.39}^{+0.35}$ & 0.57$_{-0.00}^{+0.00}$ & 2.26$_{-0.08}^{+0.08}$ & 0.000$_{-0.016}^{+0.016}$ & 0.000$_{-0.035}^{+0.033}$ & 0.02 \\
8.81--10.27 & 36 & 36 & 10 & 40.49$_{-0.55}^{+0.46}$ & 0.41$_{-0.00}^{+0.00}$ & 1.50$_{-0.05}^{+0.04}$ & 0.260$_{-0.021}^{+0.020}$ & 0.000$_{-0.031}^{+0.029}$ & 0.17 \\
	\noalign{\vskip 1mm}
      
      \hline
      \multicolumn{10}{c}{NB973 ($z=0.453$--0.499)}\\
7.24--8.24 & 13 & 13 & \ldots & 17.88$_{-0.61}^{+0.65}$ & 0.45$_{-0.01}^{+0.01}$ & \ldots & 0.072$_{-0.040}^{+0.044}$ & \ldots & \ldots \\
8.27--8.74 & 13 & 13 & \ldots & 48.96$_{-0.96}^{+1.02}$ & 0.45$_{-0.00}^{+0.00}$ & \ldots & 0.092$_{-0.015}^{+0.015}$ & \ldots & \ldots \\
8.76--8.94 & 13 & 13 & \ldots & 27.12$_{-0.95}^{+0.91}$ & 0.58$_{-0.00}^{+0.00}$ & \ldots & 0.000$_{-0.000}^{+0.000}$ & \ldots & \ldots \\
8.98--9.33 & 13 & 13 & \ldots & 34.90$_{-0.72}^{+0.65}$ & 0.52$_{-0.00}^{+0.00}$ & \ldots & 0.000$_{-0.015}^{+0.015}$ & \ldots & \ldots \\
9.53--10.43 & 13 & 13 & \ldots & 33.05$_{-1.00}^{+0.84}$ & 0.38$_{-0.01}^{+0.01}$ & \ldots & 0.406$_{-0.074}^{+0.076}$ & \ldots & \ldots \\
      \hline

    \end{tabular}
    \begin{tablenotes}
    \item \leavevmode\kern-\scriptspace\kern-\labelsep (1): Stellar mass range. (2)--(4): Number of galaxies in composite spectra for \Hg, \Hb, and \Ha. (5): Emission-line fluxes in units of 10$^{-18}$ erg s$^{-1}$ cm$^{-2}$ for \Hb. (6)--(7): \Hg/\Hb\ and \Ha/\Hb\ Balmer decrement. (8)--(9): \EBVm\ derived from \Hg/\Hb\ and \Ha/\Hb\ flux ratios. (10): \NIIl/\Ha\ flux ratio.
  \end{tablenotes}
  \end{threeparttable}
\end{table*}

\begin{table*}[ht]
  \centering
  \caption{Summary of measurements from Keck-II/DEIMOS composite spectra}
  \label{tab:keck_stacked_gals}
  \begin{threeparttable}
    \centering
    \begin{tabular}{lcccccccc}
      \hline\hline
      $\log(\Mstar/\Msun)$ & $N_{\rm H\beta}$ & $N_{\rm H\alpha}$ & $F_{\rm H\beta}$ & $F_{\rm H\alpha}$/$F_{\rm H\beta}$ & \EBVm$_{\rm H\alpha}$ & $F_{[\textsc{N~ii}]}$/$F_{\rm H\alpha}$ \\
      (1) & (2) & (3) & (4) & (5) & (6) & (7)\\
      \hline
      \multicolumn{7}{c}{NB921 ($z=0.389$--0.428)}\\
6.39--7.96 & 19 & 19 & 3.61$_{-0.14}^{+0.15}$ & 3.22$_{-0.10}^{+0.10}$ & 0.120$_{-0.032}^{+0.032}$ & 0.02 \\
7.97--8.24 & 18 & 17 & 6.92$_{-0.18}^{+0.17}$ & 2.77$_{-0.05}^{+0.05}$ & 0.000$_{-0.018}^{+0.018}$ & 0.03 \\
8.28--8.51 & 17 & 17 & 17.61$_{-0.19}^{+0.19}$ & 3.04$_{-0.02}^{+0.03}$ & 0.062$_{-0.007}^{+0.008}$ & 0.08 \\
8.53--8.99 & 18 & 18 & 28.74$_{-0.24}^{+0.23}$ & 2.87$_{-0.02}^{+0.02}$ & 0.005$_{-0.007}^{+0.007}$ & 0.12 \\
9.00--9.72 & 19 & 19 & 52.18$_{-0.49}^{+0.50}$ & 3.01$_{-0.02}^{+0.02}$ & 0.051$_{-0.008}^{+0.008}$ & 0.29 \\
	\noalign{\vskip 1mm}

      \hline
      \multicolumn{7}{c}{NB973 ($z=0.458$--0.498)}\\
7.24--8.31 & 11 & 10 & 14.56$_{-0.25}^{+0.26}$ & 2.61$_{-0.01}^{+0.01}$ & 0.000$_{-0.005}^{+0.005}$ & 0.00 \\
8.35--8.73 & 10 & 8 & 54.21$_{-0.23}^{+0.22}$ & 2.90$_{-0.00}^{+0.00}$ & 0.015$_{-0.001}^{+0.001}$ & 0.02 \\
8.76--9.05 & 11 & 9 & 40.69$_{-0.27}^{+0.26}$ & 3.15$_{-0.00}^{+0.00}$ & 0.096$_{-0.001}^{+0.001}$ & 0.03 \\
9.13--9.43 & 9 & 6 & 28.34$_{-0.38}^{+0.36}$ & 2.98$_{-0.04}^{+0.04}$ & 0.040$_{-0.014}^{+0.014}$ & 0.11 \\
9.54--10.27 & 11 & 7 & 33.62$_{-0.44}^{+0.45}$ & 2.87$_{-0.00}^{+0.00}$ & 0.003$_{-0.001}^{+0.001}$ & 0.54 \\
      \hline
      
    \end{tabular}
    \begin{tablenotes}
    \item \leavevmode\kern-\scriptspace\kern-\labelsep (1): Stellar mass range. (2)--(3): Number of galaxies in composite spectra for \Hb\ and \Ha. (4): Emission-line fluxes in units of 10$^{-18}$ erg s$^{-1}$ cm$^{-2}$ for \Hb. (5) \Ha/\Hb\ Balmer decrement. (6): \EBVm\ derived from \Ha/\Hb\ Balmer decrement. (7): \NIIl/\Ha\ flux ratio.
  \end{tablenotes}
  \end{threeparttable}
\end{table*}

\section{Summary of SFR--$\Mstar$ relation}

In this appendix, we summarise in the following table the results of our SFR-$\Mstar$ relation as function of redshift. We refer to Section~\ref{sec:mainseq} for the analyses.

\begin{landscape}
\begin{table}[ht]
  \centering
  \caption{Summary of redshift-dependent SFR--$\Mstar$ relation.}
  \label{tab:zdep}
  \begin{threeparttable}
    \centering
    \begin{tabular}{lccccccccccccccc}
      \hline\hline
      $\log\left(\frac{\Mstar}{\Msun}\right)$ & \multicolumn{3}{c}{NB704, NB711} & & \multicolumn{3}{c}{NB816} & & \multicolumn{3}{c}{NB921} & & \multicolumn{3}{c}{NB973}\\
      \cline{2-4}
      \cline{6-8}
      \cline{10-12}
      \cline{14-16}
              & $N$ & $\left< M\right>$ & $\left<{\rm SFR}\right>$ & &  $N$ & $\left< M\right>$ & $\left<{\rm SFR}\right>$ & &  $N$ & $\left< M\right>$ & $\left<{\rm SFR}\right>$  & &  $N$ & $\left< M\right>$ & $\left<{\rm SFR}\right>$\\
          (1) &     (2) & (3) & (4) 	 & &  (5) & (6) & (7)   & &  (8) & (9) & (10)  	& & (11) & (12) & (13)\\
    \hline

 6.0--6.5  & 16 & 6.30 & --2.56 (--3.10) & &  2 & 6.27 & --2.06 (--2.41) & & \ldots & \ldots & \ldots & & \ldots & \ldots & \ldots \\
 6.5--7.0  & 16 & 6.73 & --2.47 (--2.98) & & 12 & 6.78 & --1.99 (--2.31) & &   1 & 6.64 & --1.20 (--1.37) & & \ldots & \ldots & \ldots \\
 7.0--7.5  & 16 & 7.19 & --2.21 (--2.63) & & 31 & 7.32 & --1.87 (--2.15) & &  18 & 7.32 & --1.31 (--1.48) & &  3 & 7.24 & --0.52 (--0.68) \\
 7.5--8.0  & 10 & 7.75 & --1.79 (--2.05) & & 41 & 7.74 & --1.54 (--1.77) & &  63 & 7.80 & --1.31 (--1.48) & & 13 & 7.76 & --0.57 (--0.74) \\
 8.0--8.5  &  1 & 8.40 & --1.73 (--1.95) & & 41 & 8.22 & --1.32 (--1.51) & & 102 & 8.25 & --1.11 (--1.28) & & 26 & 8.26 & --0.69 (--0.85) \\
 8.5--9.0  &  1 & 8.55 & --1.53 (--1.69) & & 18 & 8.72 & --0.84 (--1.00) & &  79 & 8.73 & --0.79 (--0.95) & & 66 & 8.75 & --0.50 (--0.67) \\
 9.0--9.5  & \ldots & \ldots & \ldots & & 8 & 9.26 & --0.36 (--0.52) & & 45 & 9.23 & --0.25 (--0.41) & & 35 & 9.16 & --0.18 (--0.35) \\
 9.5--10.0 & \ldots & \ldots & \ldots & & 1 & 9.52 & --0.40 (--0.57) & & 18 & 9.77 & +0.12 (--0.05) & & 23 & 9.66 & +0.14 (--0.03) \\
10.0--10.5 & \ldots & \ldots & \ldots & & \ldots & \ldots & \ldots & & 2 & 10.25 & +0.13 (--0.04) & & 6 & 10.22 & +0.41 (+0.24) \\
	
    \hline
  \end{tabular}
  \begin{tablenotes}
    \item \leavevmode\kern-\scriptspace\kern-\labelsep (1): Stellar mass range. (2)--(13): Number of galaxies, average log stellar mass, and average log SFR (average log SFR without FUV corrections) for NB704 and/or NB711 emitters (2)--(4), NB816 emitters (5)--(7), NB921 emitters (8)--(10), and NB973 emitters (11)--(13).
    Bins with five or fewer galaxies have less reliable derived SFR measurements, possibly due to selection effects of NB emission-line flux limits at the faint end and minimum EW excess at the bright end.
  \end{tablenotes}
  \end{threeparttable}
\end{table}
\clearpage
\end{landscape}

%%%%%%%%%%%%%%%%%%%%%%%%%%%%%%%%%%%%%%%%%%%%%%%%%%

% Don't change these lines
\bsp	% typesetting comment
\label{lastpage}
\end{document}